\documentclass[sigconf]{acmart}

\AtBeginMaketitle{%

\def\EDBTISSN{xxxx-xxxx}
\def\EDBTISBN{xxx-x-xxxxx-xxx-x}
\setcopyright{none}
\copyrightyear{\copyright\ 2027 Copyright held by the owner/author(s).
  Published on xxxxx under ISBN \EDBTISBN, series ISSN \EDBTISSN.
  Distribution of this paper is permitted under the terms of the
  Creative Commons license CC-by-nc-nd 4.0}
\acmYear{2027}
\acmDOI{}
\acmISBN{}

\acmConference[EDBT '27]{Extending Database Technology}{6-9 April 2027}{Lille (France)}

\settopmatter{printacmref=false, printccs=false, printfolios=false}

\newsavebox{\ximagebox}
\newlength{\ximageheight}
\newsavebox{\xglyphbox}
\newlength{\xglyphheight}
\newcommand{\xbox}[1]%
  {\savebox{\ximagebox}{#1}%
  \settoheight{\ximageheight}{\usebox{\ximagebox}}%
  \savebox{\xglyphbox}{\color{white}\char32}%
  \settoheight{\xglyphheight}{\usebox{\xglyphbox}}%
  \raisebox{\ximageheight}[0pt][0pt]{\raisebox{-\xglyphheight}[0pt][0pt]{%
    \makebox[0pt][l]{\usebox{\xglyphbox}}}}%
    \usebox{\ximagebox}%
    \raisebox{0pt}[0pt][0pt]{\makebox[0pt][r]{\usebox{\xglyphbox}}}}

\newsavebox{\LogoBox}
\sbox{\LogoBox}{}

\fancypagestyle{firstpagestyle}{%
  \fancyhf{}%
  \fancyfoot{}%
  \fancyhead[L,C]{}
  \fancyhead[R]{}{\raisebox{15pt}[0pt][0pt]{\xbox{\usebox{\LogoBox}}}}
  }

\hypersetup{%
  pdfcopyright={CC-by-nc-nd},
  pdfcreator={LaTeX with acmart, hyperref, and EDBT modifications},
  pdfvolumenum={XX},
  pdfissuenum={X},
  pdfisbn={\EDBTISBN},
  pdfissn={\EDBTISSN}
}

}

\usepackage{amsthm}
\newtheorem{example}{Example}[section]

\usepackage[ruled,vlined]{algorithm2e}
\usepackage{graphicx}   
\usepackage{subcaption} 
\usepackage{listings}
\lstdefinestyle{promptstyle}{
    language={}, 
    basicstyle=\ttfamily\small, 
    breaklines=true, 
    columns=fullflexible, 
    frame=single, 
    frameround=tttt, 
    backgroundcolor=\color[rgb]{0.95,0.95,0.95}, 
    numbersep=5pt, 
    showstringspaces=false, 
    rulecolor=\color{black},
}

\usepackage{caption}
\DeclareCaptionType{promptblock}[Prompt Block]

\DeclareCaptionType{queryblock}[Query Block]

\usepackage{booktabs}
\usepackage{multirow}
\usepackage{balance} 

\usepackage{pdfpages}
\usepackage{xcolor}
\newcommand{\fuheng}[1]{\textcolor{black}{#1}}
\newcommand{\rev}[1]{\textcolor{black}{#1}}  

\begin{document}

\title{Access Paths for Efficient Ordering with Large Language Models}



\author{Fuheng Zhao$^1$, Jiayue Chen$^2$, Yiming Pan$^3$, Tahseen Rabbani$^2$}
\author{Sohaib$^4$, Divyakant Agrawal$^4$, Amr El Abbadi$^4$}
\author{Paritosh Aggarwal$^5$, Anupam Datta$^5$, Dimitris Tsirogiannis$^5$}
\affiliation{\institution{$^1$University of Utah \quad $^2$University of Chicago \quad $^3$UCLA \quad $^4$UC Santa Barbara \quad $^5$Snowflake}\country{USA}}

\email{fuheng.zhao@utah.edu, {jiayuechen, trabbani}@uchicago.edu, yimingpan@ucla.edu}
\email{{sohaib, divyagrawal, amr}@ucsb.edu, {paritosh.aggarwal, anupam.datta, d.tsirogiannis}@snowflake.com}


\renewcommand{\shortauthors}{First Author et al.} 

\begin{abstract}
In this work, we present the \texttt{LLM ORDER BY} semantic operator as a logical abstraction and conduct a systematic study of its physical implementations. First, we propose several improvements to existing semantic sorting algorithms and introduce a semantic-aware external merge sort algorithm. Our extensive evaluation reveals that no single implementation offers universal optimality on all datasets. From our evaluations, we observe a general scaling relationship between sorting cost and the ordering quality for LLM ranking algorithms. Building on these insights, we design a budget-aware optimizer that utilizes heuristic rules, LLM-as-Judge evaluation, and consensus aggregation to dynamically select the near-optimal access path for LLM ORDER BY. 
In our extensive evaluations, our optimizer consistently achieves ranking accuracy on par with or superior to the best static methods across benchmarks. We believe that this work provides foundational insights into the principled optimization of semantic operators essential for building robust, large-scale LLM-powered analytic systems.
\end{abstract}


\maketitle

\section{Introduction}

Recent advances in large language models (LLMs) have facilitated complex data processing, such as automatically translating natural language questions into SQL queries~\cite{zhao2024sphinteract, floratou2024nl2sql, li2023can, gao2023text, gong2025sqlens}, powering autonomous problem-solving agents~\cite{zhuge2024gptswarm, shankar2024docetl, liu2025palimpzest, zhao2023llm, sun2025agenticdata, papadopoulos2025haides, zhao2025neurdbdesignimplementationaipowered, he2025cognify}, and introducing new declarative query languages with AI operators~\cite{zeng2025adl, zhao2024hybrid,satriani2025logical,glenn2024blendsql, cortexaisql}. Building LLM declarative interfaces through semantic operators have also attracted significant attention from industry, with systems such as BigQuery ML~\cite{google_bigquery_bqml}, Snowflake AI SQL~\cite{snowflake_cortex_aisql, cortexaisql}, and Amazon Redshift's user-defined functions~\cite{aws_bedrock_redshift}.

In this work, we focus on the implementation of one specific LLM/AI semantic operator: \texttt{LLM ORDER BY}, commonly used for sorting or ranking. Sorting is a fundamental operation in data processing and underpins a wide range of analytical and decision-making tasks. \texttt{LLM ORDER BY} enables ranking items based on semantic signals, user preferences, or task-specific criteria that are difficult to capture with traditional methods. For example, in information retrieval, it is common to use LLMs to re-rank retrieved documents or passages based on the user's question or preferences~\cite{zhu2023large}. In the context of relational data processing, LLMs can be leveraged to rank selected rows based on attributes that are not explicitly stored in the relational table, such as inferred sentiment or other factual information~\cite{zhou2024serving, zhao2024hybrid}. Let's consider the following two concrete examples:

\begin{example}
A user provides a query and, after an initial index filter (e.g., BM25~\cite{robertson2009probabilistic}) on a large documents corpus, the user invokes the \texttt{LLM ORDER BY} operator to re-rank the retrieved subset based on the criteria (e.g., document relevance to \textsc{bmt medical}). This approach allows the LLM to generate high quality ordered results based on a nuanced semantic understanding of the user's query and preferences.
\end{example}

\begin{center}
\begin{minipage}{0.95\linewidth}
\label{example_passage_query}
\begin{lstlisting}[style=promptstyle]
SELECT p.id, p.text FROM (
    SELECT id, text FROM passages
    ORDER BY BM25(text, 'define bmt medical') DESC
    LIMIT 100 ) AS p
LLM_ORDER_BY(p.text, 'relevance to query: define bmt medical') DESC LIMIT 10;
\end{lstlisting}
\end{minipage}
\end{center}

\begin{example}
A company wants to rank customer feedback based on specific criteria, such as product dissatisfaction or satisfaction level, and to improve its product based on the feedback. Because "dissatisfaction" or "satisfaction" is a subjective attribute, the data analyst can invoke the \texttt{LLM ORDER BY} operator to rank the customer feedback based on this specific criteria, requiring the LLM to highlight the subtle and nuanced levels of customer sentiment. 
This capability enables the company to directly address the most critical product pain points or capitalize on successful features.
\end{example}

\begin{center}
\begin{minipage}{0.95\linewidth}
\begin{lstlisting}[style=promptstyle]
SELECT id, text FROM reviews
LLM_ORDER_BY(text, 'degree of positivity') 
DESC LIMIT 10;
\end{lstlisting}
\end{minipage}
\end{center}

Given the importance of data ordering and ranking, researchers have explored a wide range of approaches in conjunction with LLMs. Prior works mainly study three classes of methods to implement sorting on top of LLMs: (i) pointwise sorting derives the value independently for each given key (e.g., passage relevance score, review sentiment score)~\cite{drozdov2023parade, sachan2022improving, zhao2024hybrid, wang2025unify}, (ii) pairwise sorting compares two keys and then determines the relative order between them~\cite{qin2023large, luo2024prp, shah2018simple}, and (iii) listwise sorting considers an entire set of keys simultaneously to produce a ranked order~\cite{ma2023zero, sun2023chatgpt}.
There is no clear consensus on which approach is better. For example, Lotus~\cite{patel2024lotus}, a recent LLM-powered data processing system, implements LLM-based ordering using a pairwise approach with quicksort. BigQuery ML implements LLM-based ordering using a pointwise method~\cite{lao2025sembenchbenchmarksemanticquery}. Furthermore, on some benchmarks, researchers find listwise sorting often provide high accuracy~\cite{ma2023zero, pradeep2023rankvicuna}, and on other benchmarks, researchers find that pairwise sorting provide higher accuracy~\cite{qin2023large}. 




Thus, in this paper, we aim to address the following question: \textbf{\fuheng{Given a semantic LLM ORDER BY task and a fixed monetary budget, how can we maximize sorting accuracy while remaining within the budget?}} Before answering this question, we first explore whether there exists a universal optimal approach that achieves the best quality for all given tasks and datasets. To this end, we implemented existing approaches, proposed new LLM-based sorting algorithms, and conducted a systematic evaluation of different approaches within a unified experimental framework across different tasks. We observe that no algorithm outperforms the others across all tasks. In some cases, pointwise method (often criticized for its limited accuracy~\cite{qin2023large}) can, in fact, achieve strong accuracy in certain scenarios, while pairwise and listwise approaches excel in others. 
This highlights the importance of tailoring the ranking strategy to the task setting rather than relying on a one-size-fits-all solution.
This highlights the importance of tailoring the ranking strategy to the task setting rather than relying on a static \texttt{LLM Access Path}~\footnote{Inspired by access paths in relational database systems~\cite{selinger1979access}}. It also suggests that the \texttt{LLM Access Path}, i.e., the strategy for invoking an LLM to implement a semantic operator, should be carefully selected based on the query characteristics, underlying model, and data. Although this paper does not aim to optimize the selection of the access paths for all operators, we present an optimizer that determines a highly accurate sorting approach for the \texttt{LLM ORDER BY} operator.

After thoroughly examining the efficiency–effectiveness trade-offs in the initial evaluations, we conjecture a scaling relationship between compute cost and quality in semantic sorting, where increased computation consistently correlates with higher quality results. 
Building upon this insight, we propose two distinct optimizer configurations for the \texttt{LLM ORDER BY} operator: (i) an LLM-as-Judge optimizer that evaluates candidate quality, and (ii) a self-consistency optimizer that leverages aggregation to derive a robust consensus. 
The LLM-as-Judge approach leverages direct reasoning and reflection to maximize accuracy. The self-consistency optimization functions as a highly robust ensemble strategy, employing consensus aggregation to filter out anomalies and to ensure superior stability on complex ranking tasks.


To conclude, our contributions can be summarized as follows:
\begin{itemize}

    \item We propose several improvements to existing semantic sorting algorithms as well as a new semantic sorting algorithm which is an adaptation of the classical two-way merge sort algorithm~\cite{selinger1979access}.


    \item Based on empirical data, we conjecture that LLM-based sorting obeys the scaling behavior, which relates computational expenditure to output quality, forming a foundation for an intelligent optimizer.
    
    \item We design and implement an optimizer that leverages heuristics, LLM judgment, and consensus voting to achieve high ranking quality within budget constraints. 
    
    \item Across five benchmarks and three LLMs, our optimizer consistently competitive with and often exceeds the best static algorithm, whose identify changes across datasets and models.
    
\end{itemize}

This paper is structured as follows: Section 2 provides background information on semantic sorting and establishes the notation. Section 3 reviews existing algorithms and introduces our proposed extensions, including the new semantic merge sort algorithm. Section 4 details our evaluation on two benchmarks, demonstrating that no single algorithm is universally optimal and quantifying the relationship between the computational budget and accuracy. 
Section 5 introduces an optimizer that incorporates heuristic rules, LLM guidance, and consensus aggregation. Section 6 presents extensive experiments validating the effectiveness of our proposed optimizers. Finally, Section 7 concludes with a discussion of our findings.
\section{Preliminaries}  

The \textbf{LLM ORDER BY} semantic operator takes as input a list of data points along with a ranking criterion and produces an ordered list. These data points may represent diverse items, such as documents or reviews, while the ranking criterion specifies the basis for ordering, such as relevance to a user question. Unlike traditional database settings, where the ordering attribute is explicitly stored in the table, LLM-based ordering relies on semantic judgments or inferred attributes. 


Previous LLM sorting algorithms are typically designed for passage or document ranking and are commonly categorized as pointwise, pairwise, or listwise. In this paper, we adopt a slightly different terminology.
We refer to pointwise methods as \textbf{value-based}, pairwise methods as \textbf{comparison-based}, and further divide listwise methods into two variants: \textbf{external value-based} and \textbf{external comparison-based}. The intuition is that listwise methods can be viewed as batching multiple data points: either batching the derivation of values for multiple data points (external value-based) or batching the determination of the correct permutation (external comparison-based) for multiple data points. Value-based ordering methods are commonly implemented in current LLM-powered analytic systems~\cite{zhao2024hybrid, wang2025unify, lao2025sembenchbenchmarksemanticquery}, typically excelling on benchmarks that test factual accuracy (e.g., ranking based on masked attributes of sports players' height or celebrity' birthday). On the other hand, systems like Lotus~\cite{patel2024lotus} employ comparison-based ordering for semantic ordering, achieving top-tier accuracy on sentence understanding tasks.


In our work, we focus on the semantic LLM ORDER BY operator and the method of implementing the operator through standard generation APIs, i.e., treating the large language model as a semantic black box. In modern data infrastructures, models, whether proprietary or managed open-source, are predominantly exposed via high-level inference interfaces to abstract away the complexities of multi-tenant serving, hardware optimization, and security. Furthermore, emerging LLM-based analytical systems~\cite{shankar2024docetl, liu2025palimpzest, trummer2025implementingsemanticjoinoperators} increasingly rely on these standardized endpoints to ensure broad compatibility among different providers (e.g., OpenAI API~\cite{openai_structured_outputs}, Snowflake Cortex API~\cite{snowflake_cortex_aisql}, LiteLLM~\cite{BerriAI_litellm_2025}). Also, the monetary cost of using these models can be calculated directly from the input and output tokens.

\section{LLM Sorting}
\label{sec:sorting}



In this section, we review existing semantic sorting algorithms (e.g., standard pointwise~\cite{nogueira2020documentrankingpretrainedsequencetosequence}, quicksort~\cite{patel2024lotus}, and external bubble sort~\cite{sun2023chatgpt}) and present our proposed optimizations: quick sort with majority voting, and semantic external merge sort.

\subsection{Value-Based Ordering}
In value-based ordering, the LLM assigns a value to each key, and the final ranking is obtained by sorting the keys according to these values. This approach is often the default implementation in many LLM-powered data systems~\cite{zhao2024hybrid, alaparthi2025scalellm, glenn2024blendsql, lao2025sembenchbenchmarksemanticquery}. The value-based method requires $O(N)$ LLM function calls where $N$ is the number of rows. Prompt Block~\ref{lst:pointwise_prompt} shows the template used to generate the relevance score for the passage ranking task.

To even further reduce the number of LLM function calls, a commonly used technique is to use the \emph{external pointwise} approach, where $m$ keys are batched into a single prompt and the model returns a list of $m$ values~\cite{cheng2023batchpromptingefficientinference}. With the external pointwise approach, the number of LLM function calls becomes $O(N/m)$. Also, the external pointwise method often leads to less monetary cost. This cost saving is achieved by sharing the instructional prefix across a batch of keys, rather than repeating it for each individual item. 
By amortizing the prompt's token cost across $m$ items, the total number of billed input tokens is substantially reduced. 
Specifically, instead of evaluating a single key as shown in Prompt Block~\ref{lst:pointwise_prompt}, this approach provides the LLM with a list of keys and requests it to generate a corresponding list of values.

\begin{center}
\begin{minipage}{0.95\linewidth}
\captionof{promptblock}{Value-based Ranking Prompt}
\label{lst:pointwise_prompt}
\begin{lstlisting}[style=promptstyle]
Evaluate how well the passage answers the question by assigning a score from 0 to {max_point}, where 0 is not relevant and {max_point} is prefectly relevant.
Question: {question}
Passage: {key}
\end{lstlisting}
\end{minipage}
\end{center}

\begin{algorithm}[tbp]
\LinesNumbered
\caption{Quick Sort with Majority Voting}
\label{alg:llm-quicksort}
\KwIn{Dataset $D$, LLMSort$(\cdot)$, votes $v \ge 1$}

\lIf{$|D| \le 1$}{\Return $D$}

$p \gets D[0]$; $L, G, L_{initial}, G_{initial} \gets [\,]$; $\mathcal{R'} \gets \emptyset$\;
$\mathcal{R} \gets \{ (x, LLMSort(p, x)) \text{ for } x \in D[1:] \}$

\For{each $(x, r) \in \mathcal{R}$}{
    \leIf{$r = l$}{$L_{initial} \gets L_{initial} \cup \{x\}$}{$G_{initial} \gets G_{initial} \cup \{x\}$}
}

\For{each $(x, r_0) \in \mathcal{R}$}{
    $P \gets (r_0 = l) ? G_{initial} : L_{initial}$\;
    
    $Y \gets Sample(P, v-1)$\;
    $R_Y \gets \{LLMSort(x,y) \mid y \in Y \}$\;
    
    \lIf{$\texttt{all}(R_Y = l)$}{$L \gets L \cup \{x\}$}
    \lIf{$\texttt{all}(R_Y = g)$}{$G \gets G \cup \{x\}$}
    $\mathcal{R'} = \mathcal{R'} \cup (x, r_0, Y, R_Y) $  \;
}

\While{$|L| + |G| < |D|-1$}{
    $deadLock \gets \text{True}$\;
    \For{each $(x, ...) \in \mathcal{R}'$ with $x \notin L \cup G$}{
        \If{$Y \subseteq L \cup G$}{
            $deadLock \gets \text{False}$\;
            \tcp{Vote refers to Alg~\ref{alg:vote}}
            \uIf{\textsc{Vote}(tuple, L, G) = l}{
                $L \gets L \cup \{x\}$\;
            }
            \Else{
                $G \gets G \cup \{x\}$\;
            }
        }
    } 
    \If{$deadLock$}{
        \fuheng{Find any $(x, ...) \in \mathcal{R}'$ with $x \notin L \cup G$\;}
        \uIf{\textsc{Vote}(tuple, L, G) = l}{
            $L \gets L \cup \{x\}$\;
        }
        \Else{
            $G \gets G \cup \{x\}$\;
        }
    }
} 

$S_L \gets \textsc{LLMQuickSort}(L, LLMSort, v)$\; 
$S_G \gets \textsc{LLMQuickSort}(G, LLMSort, v)$\;
\Return $S_L \,\|\, [p] \,\|\, S_G$\;
\end{algorithm}

\begin{algorithm}[tbp]
\LinesNumbered
\caption{Vote Routine in Quick Sort}
\label{alg:vote}
\lIf{$r_0 = l$}{$L_{\text{vote}}, G_{\text{vote}} \gets 1.5, 0$}
\lElse{$L_{\text{vote}}, G_{\text{vote}} \gets 0, 1.5$}
\For{each peer $y_i \in Y$ with result $r_i \in R_Y $}{
    \lIf{$y_i \in L \land r_i = l$}{$L_{\text{vote}} \gets L_{\text{vote}} + 1$}
    \lIf{$y_i \in G \land r_i = g$}{$G_{\text{vote}} \gets G_{\text{vote}} + 1$}
}
\lIf{$L_{\text{vote}} > G_{\text{vote}}$}{\Return $l$}
\lElse{\Return $g$}
\end{algorithm}

\begin{algorithm}[htbp]
\caption{LLM External Merge Sort}
\label{alg:llm-external-merge-sort}
\LinesNumbered
\KwIn{Dataset $D$, batch size $m$, sorter LLMSort$(\cdot)$, two-way merger \textsc{LLMMerge}$(\cdot)$}

\textbf{Phase 1: Run generation}\;
\fuheng{Partition $D$ into $C_0,\dots,C_{\lceil |D|/m \rceil -1}$ of size at most $m$\;}
\ForPar{\fuheng{$i \gets 1$ \KwTo $\lceil |D|/m \rceil - 1$}}{
  $S_i \gets LLMSort(C_i)$ \tcp{listwise sorting}
}
\fuheng{$Runs \gets [S_0,\dots,S_{\lceil |D|/m \rceil-1}]$}\;

\BlankLine
\textbf{Phase 2: Iterative merging}\;
\While{$|Runs| > 1$}{
  $NewRuns \gets [\,]$\;
  \For{\fuheng{$i = 0, 2, 4, \dots, |Runs|-1$}}{
    \If{$i+1 \le |Runs|-1$}{
      \fuheng{$S \gets \textsc{LLMMerge}(Runs[i],\, Runs[i{+}1],\, m)$ \tcp{Algorithm~\ref{alg:llm-two-way-merge}}}
      append $S$ to $NewRuns$\;
    }
    \Else{
      append $Runs[i]$ to $NewRuns$
    }
  }
  $Runs \gets NewRuns$\;
}
\Return $Runs[1]$ \tcp{final sorted sequence}
\end{algorithm}

\subsection{Comparison-Based Ordering}

In comparison-based ordering, the LLM is asked to directly compare two or more keys and decide the ranking order. Instead of assigning independent values to each key, the model outputs a permutation of the input keys. Compared to value-based ordering, comparison-based methods are typically more expensive, as the number of comparisons can grow quadratically with the number of keys in the worst case. 
Typically, researchers have implemented quick sort and external bubble sort on top of the LLM comparators. In the case of external bubble sorting algorithm (sliding window strategy)~\cite{sun2023chatgpt}, it sorts $m$ keys at a time and then advances by $\frac{m}{2}$ steps. It allows the model to repeatedly compare overlapping groups of keys (i.e., $\frac{m}{2}$ keys), so that $\frac{m}{2}$ items of higher rank gradually 'bubble up' in successive windows.

We introduce two new comparison-based algorithms: (i) \textbf{quick sort with majority voting} to improve sorting quality compared to vanilla quick sort, and (ii) a \textbf{semantic external merge sort} inspired by the classical two-way external merge sort to reduce the cost compared to external bubble sort. Later in the evaluations, we empirically demonstrate the effectiveness of these methods, showing that our quick sort variant leads to better accuracy than vanilla quick sort, and the external merge sort yields substantial cost savings while maintaining high quality outputs.

\begin{samepage}
\begin{center} 
    \captionof{promptblock}{Comparison Prompt}
    \label{lst:pairwise_prompt}
\begin{lstlisting}[style=promptstyle]
Determine which passage answers the question better. 
Question: {question}
Passage A: {key1}
Passage B: {key2}
\end{lstlisting}
\end{center}
\end{samepage}

As shown in Algorithm~\ref{alg:llm-quicksort}, LLM quick sort algorithm follows the classical divide-and-conquer strategy, and uses an LLM comparator (LLMSort) which takes in two keys and then determines the semantic order between these two keys. Prompt Block~\ref{lst:pairwise_prompt} shows the template used to generate the correct order between two passages.
Unlike quick sort, which relies on a single comparison against the pivot to partition items, 
quick sort with majority votes aims to further improve the sorting robustness by validating the partition through majority voting. In this method, an item is compared not only with the pivot but also with a set of sampled 'peers'. The core intuition relies on transitivity: if a key $x$ is less than the pivot $p$, it should logically be less than any item $y$ that is greater than $p$. Items that achieve full consensus across all peer comparisons are partitioned immediately. Items with conflicting results are resolved iteratively; they wait until their peers are firmly classified into the Less ($L$) or Greater ($G$) sets, then a voting routine (Algorithm~\ref{alg:vote}) determines their final placement. This ensures that high-confidence decisions anchor the partition structure before uncertain cases are resolved. Only when a deadlock occurs (i.e., no remaining unresolved items have fully classified peers), then the voting routine is triggered on an unresolved item in order to resolve the cyclic dependencies.

\begin{samepage}
\begin{center} 
    \captionof{promptblock}{External Comparison Prompt}
    \label{lst:ext_comparison_prompt}
\begin{lstlisting}[style=promptstyle]
Rank the passages based on how well they answer the question in ascending order.
Question: {question}
Passages:
[1] {passage_1}
[2] {passage_2}
...
\end{lstlisting}
\end{center}
\end{samepage}

\begin{algorithm}[tbp]
\caption{LLM Two-Way Merge}
\label{alg:llm-two-way-merge}
\LinesNumbered
\KwIn{Sorted runs $L_1$, $L_2$; batch size $m$; LLMSort$(\cdot)$}
\KwOut{Merged sorted run $S$}

$i \gets 0,\; j \gets 0$ \tcp{read positions in $L_1$ and $L_2$}
$S \gets [\,]$\;
$h \gets \max(\lfloor m/2 \rfloor, 1)$\;

\While{$i < |L_1|$ \textbf{or} $j < |L_2|$}{
  \If{$i \ge |L_1| \;\vee\; j \ge |L_2|$}{
    append $\big(i \ge |L_1| \;?\; L_2[j{:}] \;:\; L_1[i{:}]\big)$ to $S$\;
    \textbf{break}\;
  }

  \tcp{Buffer up to $h$ items from each run}
  $a \gets \min(h,\, |L_1| - i)$;\quad $b \gets \min(h,\, |L_2| - j)$\;
  $B \gets L_1[i{:}i{+}a] \,\cup\, L_2[j{:}j{+}b]$\;
  $R \gets LLMSort(B)$\;

  \tcp{Emit in sorted order until one run's buffer is exhausted}
  $e_1 \gets 0,\; e_2 \gets 0$\;
  \For{each $x \in R$}{
    append $x$ to $S$\;
    \lIf{$x \in L_1$}{$e_1 \gets e_1 + 1$}\lElse{$e_2 \gets e_2 + 1$}
    \lIf{$e_1 = a$ \textbf{or} $e_2 = b$}{\textbf{break}}
  }
  $i \gets i + e_1$;\quad $j \gets j + e_2$\;
}
\Return $S$
\end{algorithm}

To address the limitation of high running cost in the external bubble sort proposed in RankGPT~\cite{sun2023chatgpt}, we propose a new semantic external merge sort approach inspired by the relational external merge sort approach~\cite{knuth1997taocp}. As shown in Algorithm~\ref{alg:llm-external-merge-sort}, our approach operates in two distinct phases: First, the Run Generation phase divides the input into sorted runs of size $m$, using \textsc{LLMSort} with Prompt Block~\ref{lst:ext_comparison_prompt}. Second, the Merge Phase (Algorithm~\ref{alg:llm-two-way-merge}) iteratively merges two sorted runs into one sorted run. The two-way merge operation employs a sliding-window mechanism: it buffers $m$ keys ($m/2$ keys from each sorted run) for the \textsc{LLMSort} to derive a partial ordering on this buffer. From this local ranking, we extract items and add them to the final output until the candidates from one of the input runs are exhausted within the current window. 
This buffer must be refilled again before the next extraction, as the unseen element from the exhausted run must be compared with the remaining items in the buffer to determine the total order. This process continues until one of the input runs is completely exhausted.

\subsection{Limit K}
\label{sec:limitk}

A common usage pattern is to combine the \textbf{ORDER BY} clause with \textbf{LIMIT K}, where the user wants to retrieve $K$ items from the sorted results. This pattern offers an additional opportunity to save the number of LLM invocations and costs. 
\fuheng{For example, the external bubble sort~\cite{sun2023chatgpt} issues $O(\frac{2N}{m})$ LLM
calls per round and places $\frac{m}{2}$ keys in their positions per round; retrieving the top $K$ keys
therefore needs $\frac{2K}{m}$ rounds, reducing the cost from $O(N^2/m^2)$ (full sort) to $O(KN/m^2)$.}
See Table~\ref{tab:complexity} for the number of LLM function calls used in each algorithm in both full sort and LIMIT $K$. Assume the user wants to retrieve the first $K$ rows in descending order based on some criteria (e.g, retrieve the $K$ most relevant passages to the user question). In quick sort, after it partitions rows into $L$ (lesser) and $G$ (greater) lists comparing each row to the pivot and $v-1$ validation samples, if $G$ contains more than $K$ rows, then only $G$ needs to be sorted and $L$ can be discarded. This is also known as partial quick sort~\cite{Martinez2004partial}. It can be seen as a quick select, which uses $O(N)$ comparisons to find the partition with $K$ rows, and then applies quick sort on $K$ rows. For External Merge Sort (Algorithm~\ref{alg:llm-external-merge-sort} and Algorithm~\ref{alg:llm-two-way-merge}), it first performs the run generation phase (i.e., generating sorted runs with $m$ rows) and then performs 2-way merge. \fuheng{The run generation phase requires $\frac{N}{m}$ LLM calls.} 
During the 2-way merge, if the merged list reaches $K$ items, then it can discard all the rest. As a result, having the LIMIT $K$ clause also reduces the number of invocations for external merge sort. For the first $O(log\frac{K}{m})$ rounds, each run size will be less than $K$, and each round requires $O(\frac{N}{m})$ function calls. 
Once the run size reaches $K$, the behavior changes. For all subsequent rounds, we merge two runs of size $K$ but strictly cap the output at $K$ items (discarding the remaining $K$). As a result, the total number of items decrease by a factor of 2 for each round. Hence, the LLM function calls decrease from $\frac{N}{m}$, to $\frac{N}{2m}$, to $\frac{N}{4m}$, etc. This is a geometric sequence which is upper bounded by $O(\frac{N}{m})$. \fuheng{The Merge phase requires $O(\frac{N}{m}(1+logK))$}. Hence, the total LLM function calls can be approximated as $O(\frac{N}{m}(2+log\frac{K}{m}))$.



\begin{table}[]
    \centering
    \begin{tabular}{l c c} 
        \toprule 
        & \multicolumn{2}{c}{LLM Calls} \\ 
        \cmidrule(lr){2-3} 
        Algorithm & Full Sort     & Limit $K$ \\ 
        \midrule 
        Pointwise  & $O(N)$    & $O(N)$        \\
        Ext Pointwise & $O(N/m)$        &  $O(N/m)$        \\
        Quick Sort ($v$)  & $O(vNlogN)$  & $O(v (N + KlogK))$ \\
        Ext Bubble & $O(N^2/m^2)$ & $O(KN/m^2)$ \\
        Ext Merge & $O(\frac{N}{m}(1+log\frac{N}{m}))$ & $O(\frac{N}{m}(2+log\frac{K}{m}))$ \\
        \bottomrule 
    \end{tabular}
    \caption{\fuheng{LLM call complexity by algorithm, where $N$ is the number of keys, $m$ is batch size, $v$ is votes, and top-$k$ items.}}
    \label{tab:complexity}
\end{table}
\section{Initial Evaluation}
~\label{sec:observation}

\begin{figure*}[htbp]
    \centering

    \begin{tabular}{@{}c@{\hspace{1em}}l@{}}
        
        \rotatebox{90}{NBA} &

        \begin{minipage}{0.96\textwidth} 
            \begin{subfigure}{0.32\linewidth} 
                \centering
                \includegraphics[width=\linewidth]{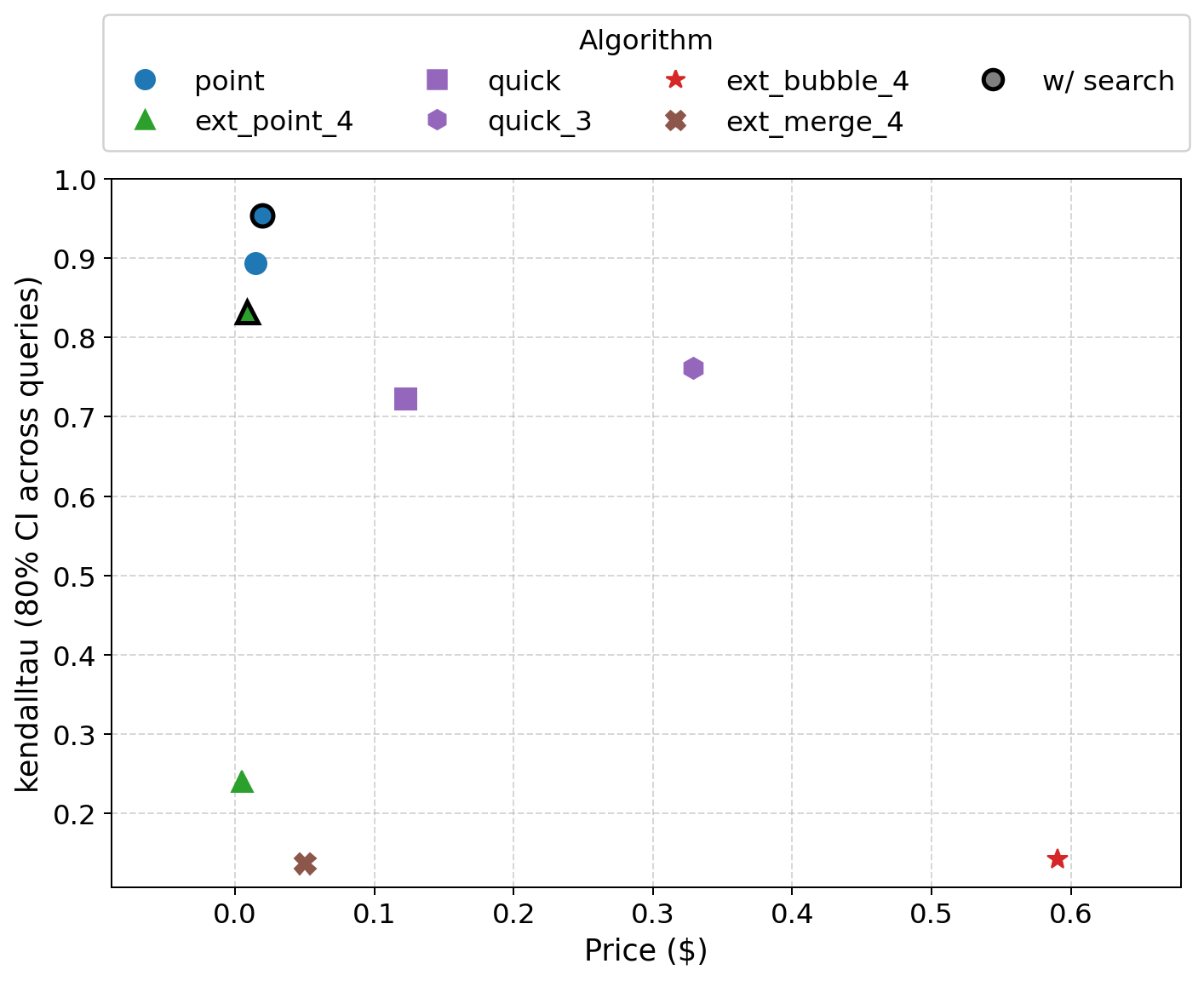}
                \caption{\fuheng{llama-3.1 70b}}
                \label{fig:sub1}
            \end{subfigure}
            \hfill
            \begin{subfigure}{0.32\linewidth} 
                \centering
                \includegraphics[width=\linewidth]{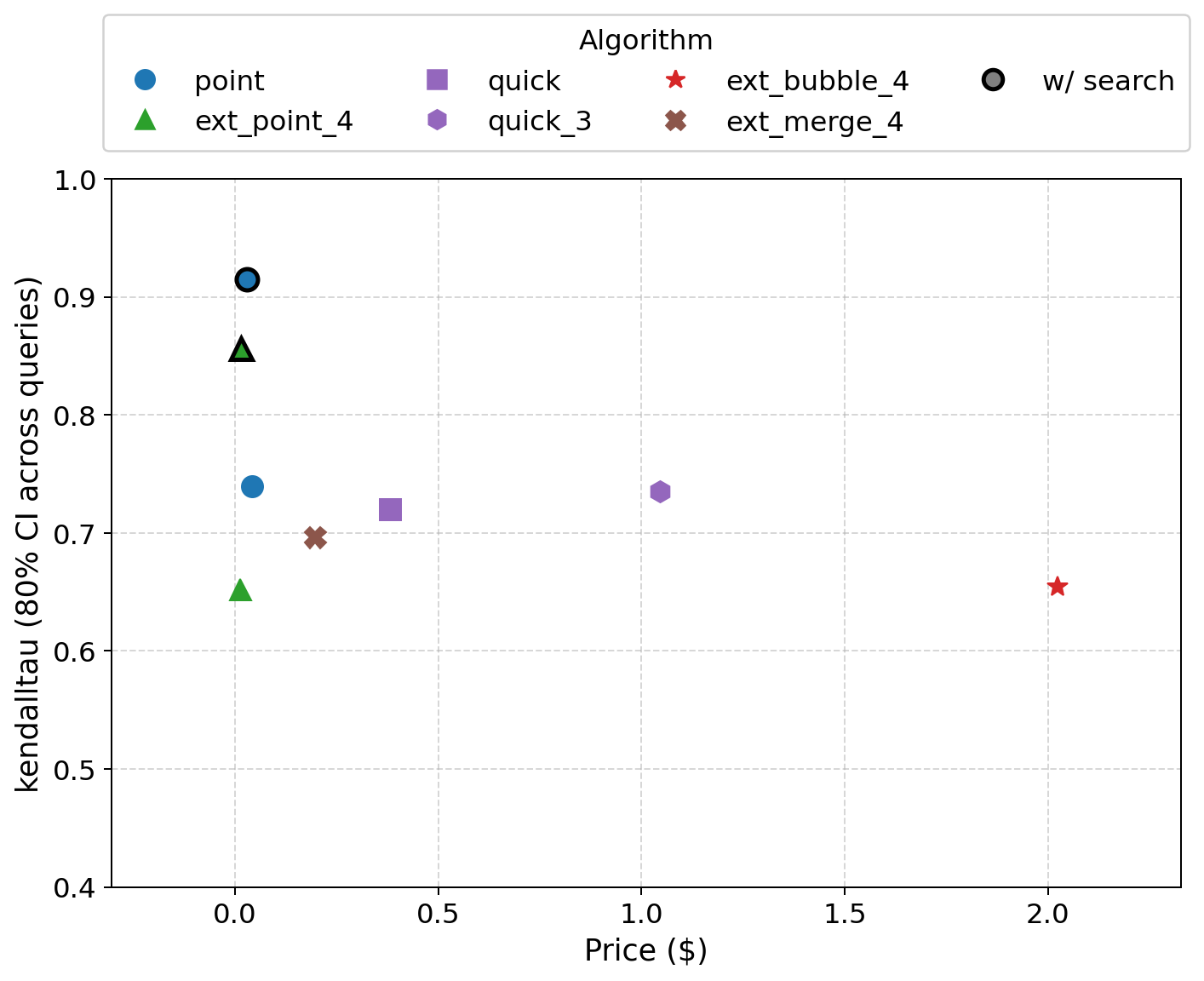}
                \caption{\fuheng{GPT-5 mini}}
                \label{fig:sub2}
            \end{subfigure}
            \hfill
            \begin{subfigure}{0.32\linewidth} 
                \centering
                \includegraphics[width=\linewidth]{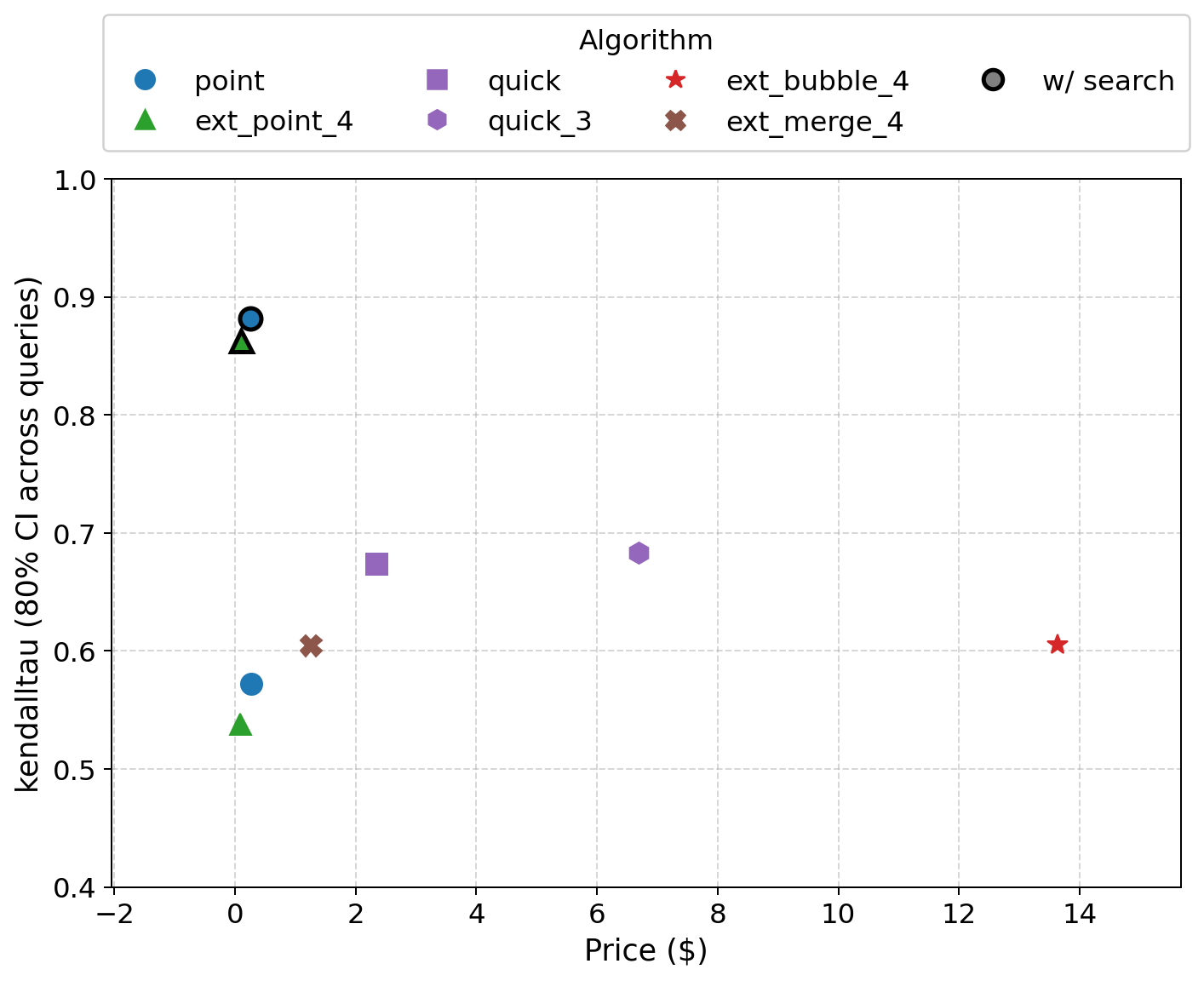}
                \caption{\fuheng{Haiku-4.5}}
                \label{fig:sub3}
            \end{subfigure}
        \end{minipage}
    \end{tabular}

    \begin{tabular}{@{}c@{\hspace{1em}}l@{}}
        
        \rotatebox{90}{DL19} &

        \begin{minipage}{0.96\textwidth} 
            \begin{subfigure}{0.32\linewidth} 
                \centering
                \includegraphics[width=\linewidth]{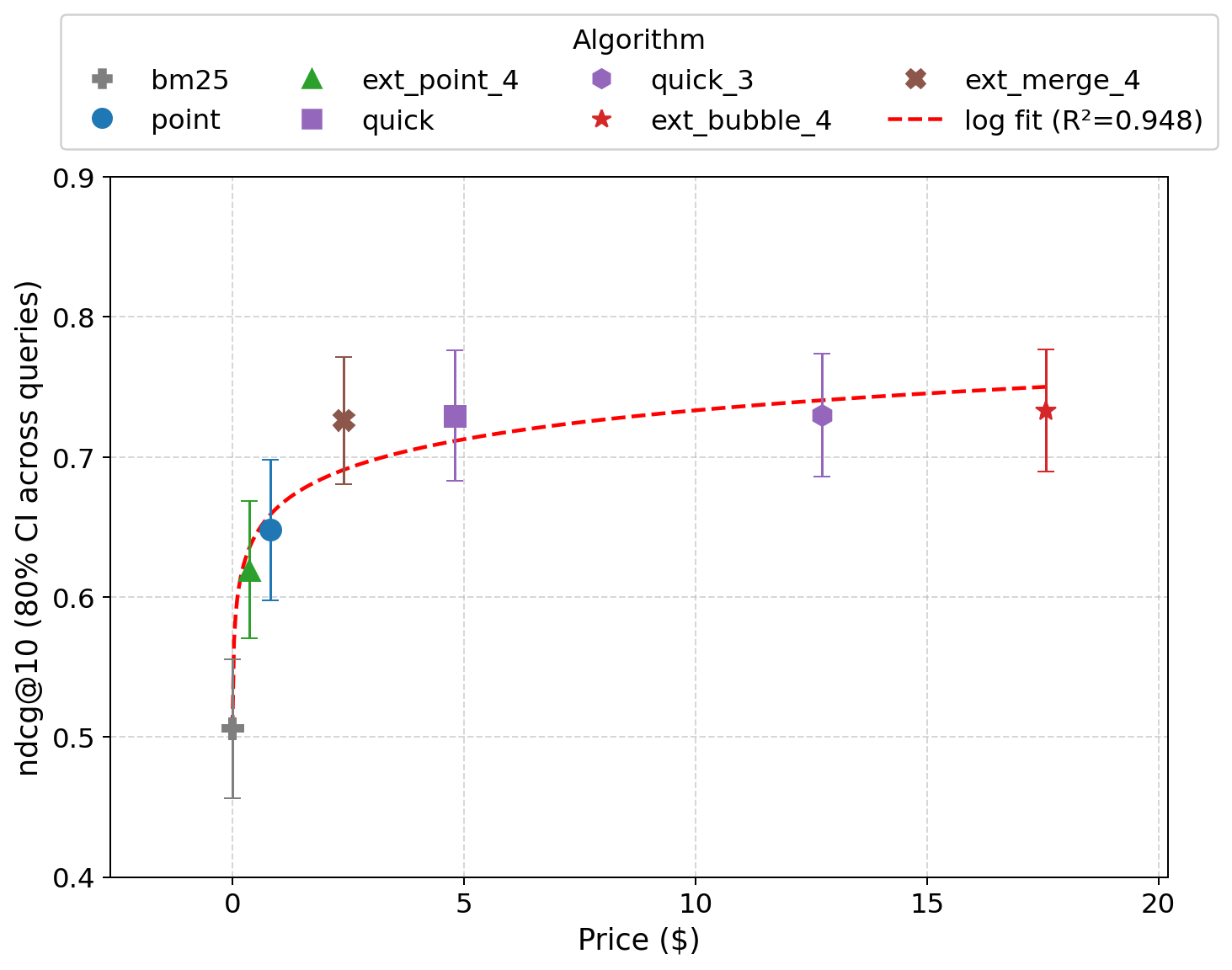}
                \caption{\fuheng{llama-3.1 70b}}
                \label{fig:sub5}
            \end{subfigure}
            \hfill
            \begin{subfigure}{0.32\linewidth} 
                \centering
                \includegraphics[width=\linewidth]{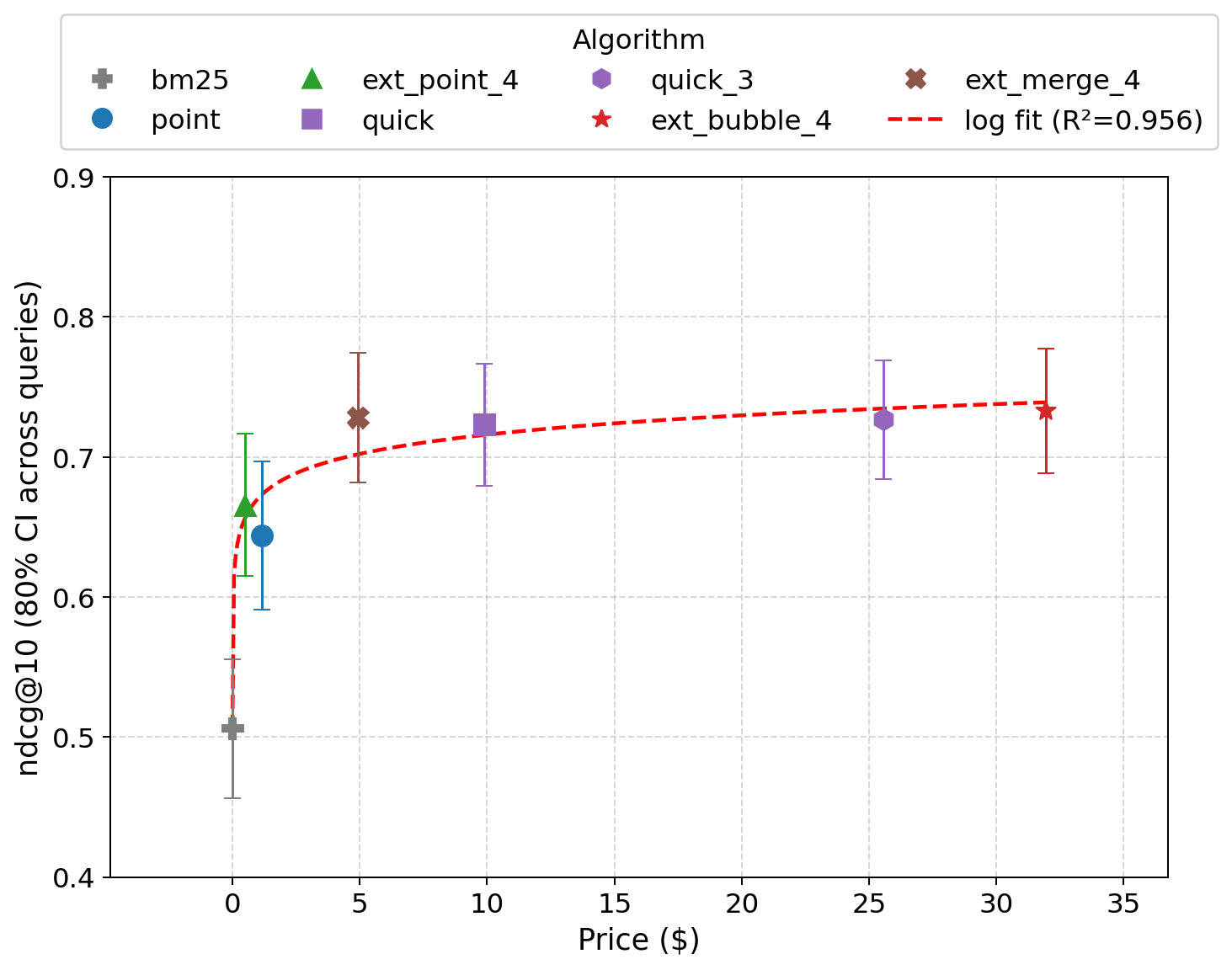}
                \caption{\fuheng{GPT-5 mini}}
                \label{fig:sub6}
            \end{subfigure}
            \hfill
            \begin{subfigure}{0.32\linewidth} 
                \centering
                \includegraphics[width=\linewidth]{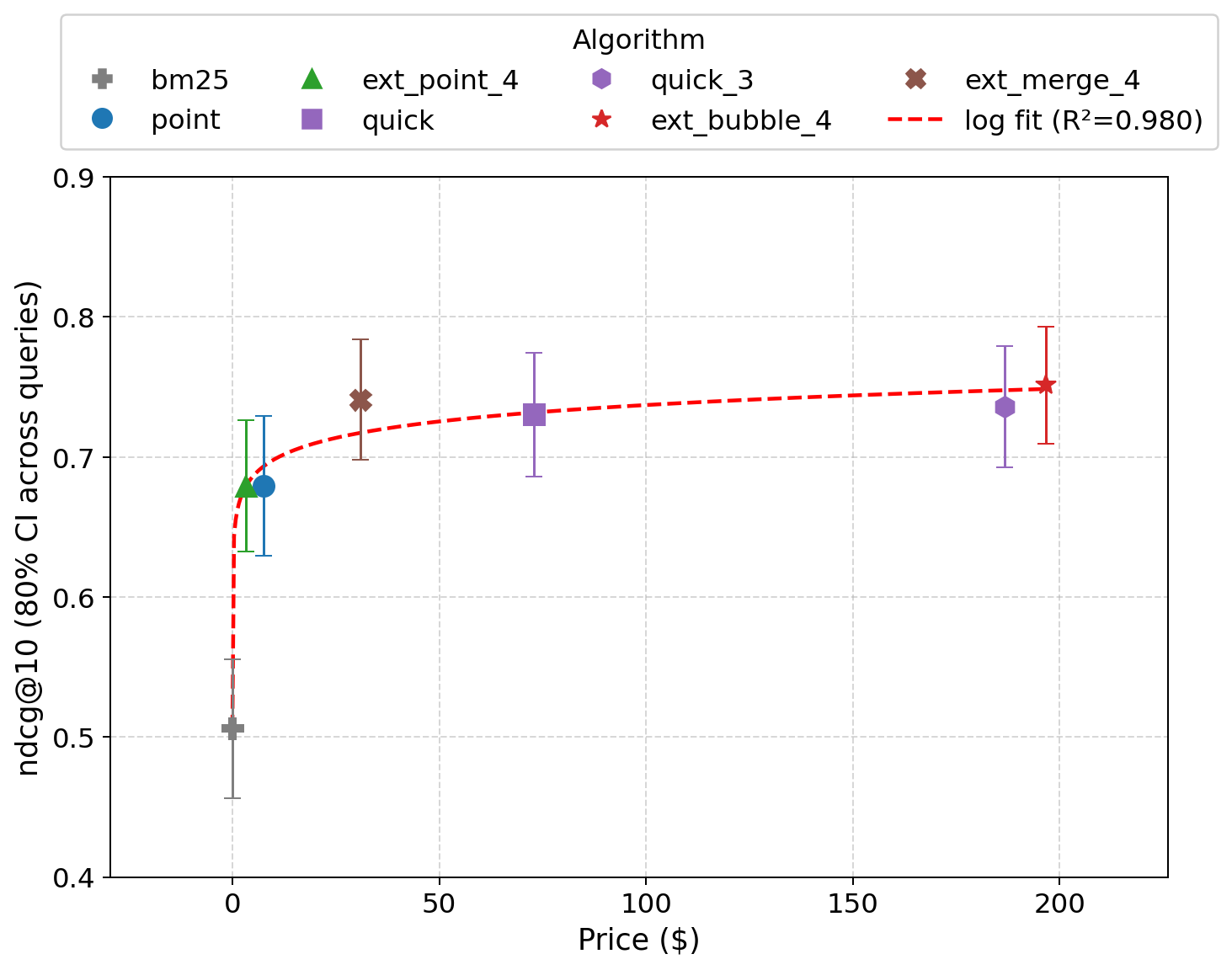}
                \caption{\fuheng{Haiku-4.5}}
                \label{fig:sub7}
            \end{subfigure}
        \end{minipage}
    \end{tabular}

    \caption{\fuheng{Sorting accuracy vs. the monetary budget (\$). While more budget generally leads to better accuracy, these algorithms often display a log-linear growth pattern (red curve), with accuracy increasing quickly at first before plateauing.}}
    \label{fig:tradeoff}
\end{figure*}

In this section, we evaluate our proposed algorithms alongside existing baselines on two different datasets. We report results in terms of both effectiveness (ranking quality) and cost (price), highlighting the trade-offs between value-based and comparison-based approaches. The main question we would like to answer in this section: \textbf{Is there a single sorting algorithm that always gives the best ranking quality?}

\noindent \textbf{Datasets.} Two datasets were used:
\begin{itemize}
    \item NBA: Contains 200 player names~\cite{openintro_nba_heights} ordered by height.

    \item DL19: TREC-DL19 is a widely used benchmark dataset in information retrieval literature~\cite{DL19}. The test sets contains 43 queries and 8.8 million passages.
\end{itemize}

\subsection{Implementation and Metrics}
\fuheng{In our experiments, all LLM models (Llama3.1-70b, OpenAI GPT-5 mini and Claude Haiku-4.5) are directly accessed through Snowflake Cortex REST API.
For all experiments, we use the structured API and set the temperature to 0, which adopted greedy decoding to obtain the most deterministic outputs~\cite{openai_faq_temperature}, and set reasoning effort to minimal.} 
The input JSON schema was designed following the best practices of chain-of-thought prompting~\cite{wei2022chain}, as recommended in the OpenAI structured outputs guide~\cite{openai_structured_outputs}. 
All algorithms were implemented in Python. The pointwise~\cite{liang2022holistic} ($point$) and external pointwise ($ext\_point$) methods use the rating scale prompt from~\cite{Zhuang2024BeyondYesAndNo}; comparison based algorithms follow the prompt design in~\cite{qin2023large, ma2023zero}; and external comparison based algorithms utilize the listwise prompt from~\cite{sun2023chatgpt}.

Notably, as we later demonstrate, the value-based approaches achieve high accuracy on strictly factual, non-reasoning tasks, such as estimating the heights of NBA players. Such high accuracy is achieved because these queries directly leverage the extensive world knowledge embedded within the models' pre-training corpora~\cite{morris2025languagemodelsmemorize, noroozizadeh2025deepsequencemodelstend}.
To maximize the factual accuracy in these scenarios, we naturally augmented the value-based approaches with a web search tool, allowing the model to ground its values based on web search information.

In evaluating sorted-ness on the NBA dataset, we adopt the well-established kendall’s tau metric~\cite{kendall1938new}, with values near 1 indicating strong agreement and values near -1 indicating strong disagreement. In evaluating the sorted-ness on DL19, we adopt the standard \textit{nDCG@10} metric~\cite{wang2013theoretical}. In addition, we follow the conventions in DL19, in which the LLM Order By is applied on the top 100 passages retrieved by BM25 for each query~\cite{robertson2009probabilistic}. We directly use the \texttt{pyserini} toolkit~\cite{lin2021pyserini} for BM25 retrieval. Normalized discounted cumulative gain (nDCG) yields values from 0.0 to 1.0, where higher values indicate rankings that are closer to the ideal order.

\subsection{Observations}

\noindent \textbf{NBA Player Height Ranking.} The first row of graphs in Figure~\ref{fig:tradeoff}, where x-axis is monetary budget and y-axis reports Kendall's tau score comparing predicted orderings with the ground truth for the NBA dataset. Overall, the \emph{value-based} approach emerges as the top performing algorithm among all three models. 
The pointwise method consistently achieves high accuracy across all models, whereas the performance of the external pointwise method is slightly lower. Furthermore, when equipped with web search capabilities, the pointwise approach can achieve near-perfect accuracy. Sometimes, web search can return conflicting information, typically due to discrepancies in how players' heights are recorded (e.g., measured with shoes v.s. without shoes).
The success of the value-based method implies that for factual data, LLMs excel at directly inferring statistics they have memorized from their training corpora. In contrast, comparison-based approaches consistently underperform. Having the chain-of-thought reasoning steps in comparison based approaches does not close this gap, suggesting that the direct retrieval of internalized facts is more reliable than multi-step reasoning. 
\fuheng{Additionally, our proposed quick sort with majority voting ($quick\_3$) always outperforms standard quick sort in accuracy across all three models, albeit at a higher cost.} Our proposed semantic external merge sort algorithm achieves performance comparable to semantic external bubble sort, while significantly reducing the overall monetary costs.

\noindent \textbf{DL19 Passage Ranking.}
The second row of graphs in Figure~\ref{fig:tradeoff} reports the evaluation of different algorithms in DL19. The x-axis is the dollar cost for running each access paths and the y-axis is the mean $nDCG@10$ across 43 queries (higher is better). 
\fuheng{The error bars denote the $80\%$ confidence interval of the per-query $nDCG@10$, and
the dashed logarithmic fit achieves $R^2 \ge 0.95$ across all three models, indicating that quality improves log-linearly with cost with strongly diminishing returns.}
For passage ranking tasks, comparison-based sorting algorithms excel. Our proposed semantic external merge sort ($ext\_merge\_4$) and quick sort with majority voting ($quick\_3$) consistently achieve the best mean $nDCG@10$. \fuheng{For example, using GPT-5 mini, standard quick sort and external bubble sort achieve average scores of 0.72 and 0.73, respectively. Our proposed quick sort with majority voting and external merge Sort both attain the score of 0.73. These performance gains come with distinct cost trade-offs: the additional voting mechanism increases the cost of quick sort with majority voting by nearly 2.5$\times$ compared to the standard variant. Conversely, external merge sort maintains high accuracy while decreasing costs by 6$\times$ relative to external bubble Sort.}
The value based approach performs worse for passage ranking. When the task is not centered on deriving factual information, value-based approaches are less effective, as they lack the comparative reasoning needed to capture subtle relevance signals, which has also been observed in prior works~\cite{ma2023zero, pradeep2023rankvicuna}. 


\noindent \textbf{Trade-off between Sorting Accuracy and Computation.}
\fuheng{Our initial observations reveal a consistent trade-off between sorting accuracy and computational cost in semantic ranking tasks (e.g., DL19). In relational \textsc{Order By}, any correct sorting algorithm produces the same deterministic output, so the choice of algorithm is purely a matter of efficiency. In a \emph{semantic} \textsc{Order By}, in contrast, the algorithm directly determines output quality with different LLM access paths. As shown by the DL19 results in Figure~\ref{fig:tradeoff}, sorting accuracy ($nDCG@10$) and monetary cost are tightly coupled: the across-query means follow a log-linear trend, with a high coefficient of determination ($R^2$ between $0.95$ and $0.98$ across the three models). This trend reflects diminishing returns: accuracy improves rapidly with modest initial spending but plateaus as cost continues to grow. While the $80\%$ confidence intervals show that individual queries vary, the mean trend across the 43 queries remains stable and closely tracks the logarithmic fit. The overarching pattern is clear: higher computational expenditure generally yields higher sorting accuracy.}

\begin{figure}
    \centering
    \includegraphics[width=0.9\linewidth]{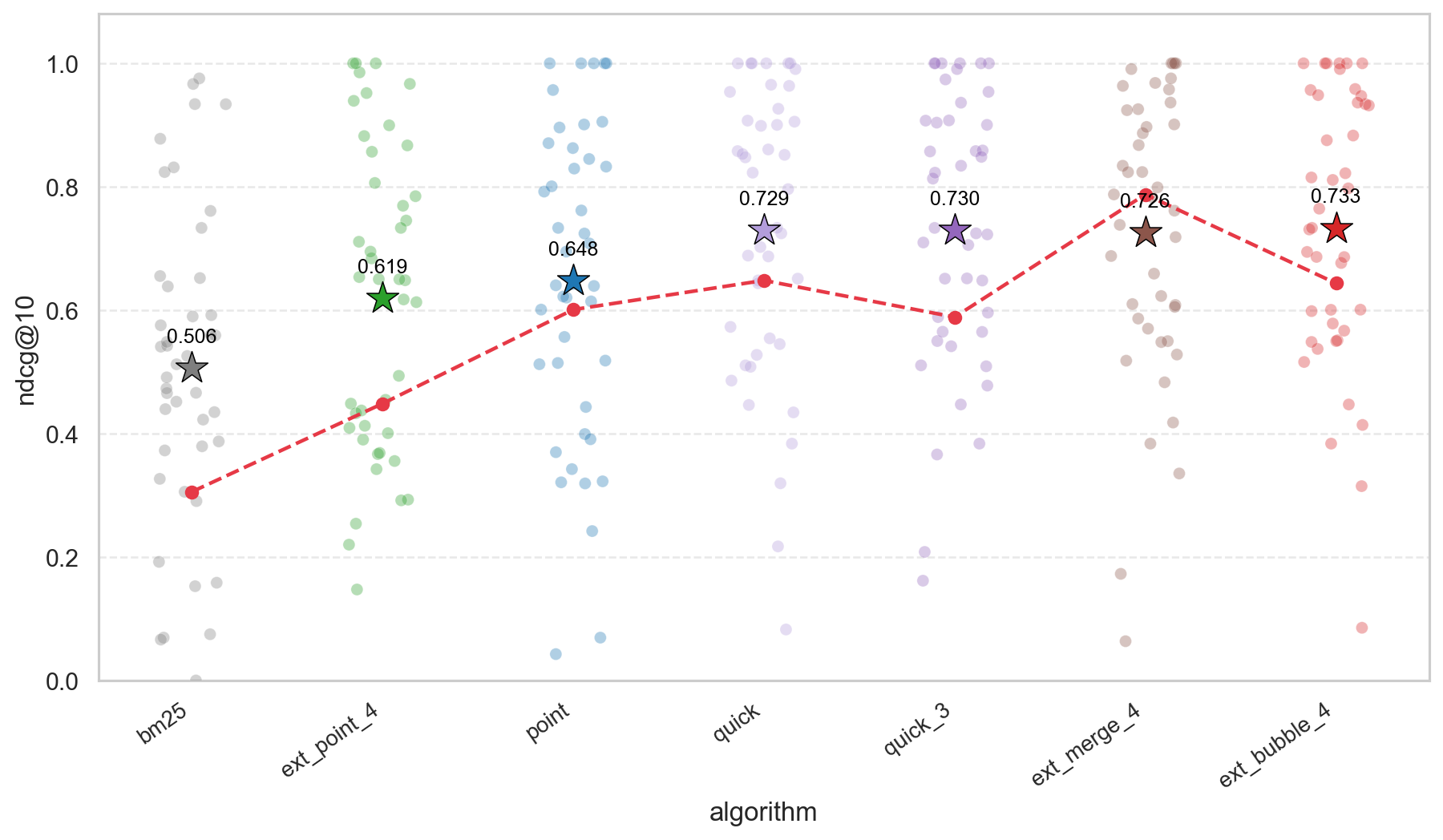}

    \caption{\fuheng{Accuracy distribution across algorithms for llama-3.1 70b.} Each dot represents a query; star markers denote the mean accuracy.}
    \label{fig:DL19llama70bPoitnCloud}
\end{figure}

\noindent \textbf{A universal best algorithm?}
Our results demonstrate that no single sorting algorithm is universally optimal across different domains. For instance, when ranking factual data such as NBA player heights, the pointwise approach excels and, with web search tools, it can reach nearly perfect accuracy. On the other hand, comparison-based approaches excel for reasoning-heavy tasks like DL19 passage relevance. To better understand this variance in reasoning tasks, we further analyzed the 43 individual user queries contained within the DL19 dataset.

Figure~\ref{fig:DL19llama70bPoitnCloud} demonstrate the accuracy distribution for each algorithm for the llama-3.1 70b model. The star denotes the mean $nDCG@10$ for each algorithm. The x-axis represents the different algorithms and the y-axis is $nDCG@10$. 
The sorting accuracy can change drastically by using different algorithms. 
The dashed line tracks the accuracy for the specific query with id 1113437. 
\fuheng{It was chosen because it demonstrates a case where both quick sort has better ranking quality than quick sort with majority vote and external merge sort has better ranking quality than external bubble sort, despite quick sort and external merge sort having lower mean $nDCG@10$ than quick sort with majority vote and external bubble sort respectively.}
This example illustrates that within the same passage ranking benchmark, there is still no universally optimal algorithm across all queries.

\section{Optimizer Design}
\label{sec:optimizer}

\fuheng{Designing an efficient optimizer for selecting the physical implementation of the semantic LLM Order By operator poses two challenges: (i) understanding the ranking quality of each candidate access path without
any ground-truth labels; and (ii) deciding which
access path(s) to execute from a small sample of the input, rather than running every candidate over the entire dataset.}
We look back at classical database query optimization literature~\cite{wu2016samplingbasedqueryreoptimization, Gibbons1998, Chen2006}. A recurring trend is the reliance on a subset of data samples or data summaries to inform algorithmic decision-making.
Taking inspirations from the classical query optimizations, we build our optimizer by first sampling a subset of the data. Then, we run several candidate algorithms in parallel, analyze the resulting performance (e.g., monetary cost and accuracy) metrics, and determine the near-optimal configuration. 



\subsection{Cost Estimation}

Before diving into the design of our optimizer, we first establish a methodology for monetary cost estimation. Unlike traditional database queries where the result is always 100\% accurate, AI SQL queries have inherent quality and cost trade-offs. Often the user wants to obtain a high quality result constrained by a cost budget. Hence, our optimizer needs to select the algorithms with respect to the user's cost constraint.

Estimating the monetary cost for each algorithm with fixed parameters is relatively straightforward. In Table~\ref{tab:complexity}, we list the number of LLM calls for each algorithm based on the number of input data points and algorithm parameters. After the sampling step, several algorithms are executed in parallel resulting in their running costs. 
This empirical measurement is crucial, as it provides the actual running cost to calibrate the theoretical cost models. 
Specifically, we can directly scale the observed cost per LLM call across the different algorithms and input sizes to accurately project the total expenditure for the original datasets. By the law of large numbers, the higher sample ratio will lead to lower estimation error. In our experimental section, we show that a small sample size can already provide quite accurate monetary cost estimation.

\begin{example}
The cost of running the pointwise LLM based sorting on a sample of size 20 is $\$0.02$. With thelinear behavior, our estimation of running the pointwise LLM sorting over 1000 data points will be $\$0.02 \cdot \frac{1000}{20} = \$1$.
\end{example}

\begin{example}
The cost of running the external bubble sort algorithm with batch size of four on the sample of size 20 is $\$0.05$. 
Since bubble sort has quadratic bevavor,  our estimation on running the external bubble sort algorithm with batch size of four over the 1000 data points will be $\$0.05 \cdot \frac{1000^2/4^2}{20^2/4^2}= \$125$.
\end{example}

\subsection{Factual World Knowledge}
\label{subsec:optimization_rule}

In Section~\ref{sec:observation}, we showed that for ordering based on factual world knowledge tasks, LLMs can directly recall factual data. The memorization of training corpora is an inherent characteristic of large language models and is essential for emergent reasoning abilities~\cite{Brown_2021,morris2025languagemodelsmemorize, wang2025generalizationvsmemorizationtracing}. The value-based methods achieve high accuracy with minimal cost. Furthermore, as demonstrated in Section~\ref{sec:observation}, pairing the pointwise approach with web search capabilities can achieve nearly perfect accuracy, grounding LLM results based on online sources.


Motivated by this observation, the challenge for ordering based on factual world knowledge becomes: \textit{Can we successfully identify whether the information needed to order items is world knowledge?} To address this question, we follow the methodology established in the field of membership inference in LLMs, and adopt the Inquiry Prompt from~\cite{Wen2024Membership} as shown in Prompt~\ref{lst:inquiry_prompt}, which achieves high membership inference accuracy and relies only on the generation REST API. In the evaluations, we find that the Inquiry Prompt effectively distinguishes between world knowledge and unknown entities.

\begin{center} 
    \captionof{promptblock}{Inquiry Prompt}
    \label{lst:inquiry_prompt}
\begin{lstlisting}[style=promptstyle]
Can the following ranking task be solved using factual world knowledge?
Task: {description}
Examples: {examples}
\end{lstlisting}
\end{center}

We run the inquiry on a sample of the input data and if the LLM identified that this ranking task can be solved based on factual knowledge, the optimizer automatically uses \textbf{pointwise} with web search ranking strategy. 
If the LLM classifies the task as subjective or reasoning-dependent, the system safely falls back to the LLM-as-Judge or consensus aggregation approaches described later in this section.
While the pointwise method is highly effective for factual sorting, its utility in broader semantic ordering is constrained by a lack of relative comparative context. By classifying the query upfront, we ensure that this method is deployed specifically when the model can function as a reliable knowledge retriever.

\subsection{Candidate Algorithms}
\fuheng{For LLM \textsc{Order By}, the initial candidate access paths include all value-based and comparison-based algorithms. Several candidates are parameterized: the external bubble and external merge sorts depend on a \textit{batch size} $m$ (the number of keys packed into a single LLM comparison call), while quick sort with majority voting depends on an integer \textit{vote} count $v$ (the number of independent comparisons aggregated per partitioning decision); each hyperparameter setting yields a distinct candidate. Because the number of valid configurations is large, bounded only by the model's context window and the budget, empirically evaluating every configuration over the entire dataset is prohibitively expensive.}

We therefore appeal to our observation from Section~\ref{sec:observation} that \textbf{higher computation generally yields higher accuracy}. This principle justifies fully utilizing the available budget, favoring the more thorough access paths the budget can afford rather than defaulting to the cheapest path.
To do so without executing every candidate in full, we estimate each candidate's price from a small sample: we run it on a sample subset, measure the monetary cost per LLM invocation, and extrapolate to the full input size using the LLM-call complexity in Table~\ref{tab:complexity}. Candidates whose projected cost exceeds the user-specified budget are discarded, and the surviving within-budget candidates form the pool from which the final choice is made. This design is flexible and extensible: incorporating a new semantic sorting
algorithm merely requires specifying its LLM-call cost function
(Table~\ref{tab:complexity}) to immediately include it in the competitive
candidate pool.

\noindent \textbf{The main challenge.}
\fuheng{
Given a pool of candidate algorithms that satisfy the user's cost budget, the remaining task is to select a subset that
produces the highest-quality ranking. Each selected algorithm is executed over the full input exactly once. The core difficulty is that we must assess ranking quality \emph{without} ground-truth labels. We address this with two label-free methodologies, detailed in the next two subsections: an \emph{LLM-as-a-judge} approach that selects a single access path,
and a \emph{consensus} approach based on reciprocal rank fusion (RRF)~\cite{RRF} that selects
and aggregates an ensemble of access paths.}

\subsubsection{LLM-as-Judge Optimizer}

A common approach in evaluating complex tasks is the emergence of "LLM-as-a-Judge", where LLMs are employed as the evaluator. LLM-as-a-Judge approach has been adopted in many different domains~\cite{gu2025surveyllmasajudge}. In the context of databases, researchers have used LLM-as-a-Judge to evaluate LLM powered natural language to SQL query translation~\cite{zhao2023llm, kim2024flexexpertlevelfalselessexecution} and to evaluate LLM generated candidate query plans~\cite{shankar2024docetl, ji2025table}. We explore whether we can also leverage an LLM to determine the best algorithm for a sorting task. We run all candidate algorithms on the sampled data. The resulting list of output rankings, each tagged with a unique candidate identifier, is then compiled and presented. The LLM Judge is subsequently tasked with evaluating this aggregate output and identifying the candidate identifier corresponding to the optimally sorted result for the given input. The list-wise Prompt~\ref{lst:judgeprompt} shows the template used to generate the LLM Judge decision, including the task instruction, the sampled keys, and the execution results of the candidate algorithms. This chain-of-thought~\cite{wei2022chain} prompt is designed in accordance with industry-standard evaluation frameworks such as TruLens~\cite{trulens}.

\begin{samepage}
\begin{center} 
    \captionof{promptblock}{LLM-as-a-Judge Prompt}
    \label{lst:judgeprompt}
\begin{lstlisting}[style=promptstyle]
You are an EXPERT RANKED RESULT RATER. You are given a ranking criteria and a list of CANDIDATE RANKED RESULTS
Your task is to find the best ranked result which sort the input keys based on the ranking criteria.
keys: {keys}
Ranking criteria: {criteria}
CANDIDATE RANKED RESULTS:
{rankings}

You should think step by step about the ranking and return the integer identifier of the best ranking.
\end{lstlisting}
\end{center}
\end{samepage}

\subsubsection{Self-Consistency Optimizer}

\fuheng{Our second quality estimator is a heuristic \emph{self-consistency} optimizer, draws on voting theory~\cite{Saari2023SelectingBorda, heilman2022noisestabilityrankedchoice}, and the LLM self-consistency paradigm~\cite{wang2023selfconsistencyimproveschainthought}. The intuition is that while any single sorted run may contain isolated reasoning errors or hallucinations, the consensus of multiple runs is statistically more likely to reflect the true semantic order. We therefore treat the ranking produced by each candidate algorithm as a set of expressed preferences and fuse these ``votes'' into a consensus ranking that serves as a label-free proxy for the ground truth.}

\fuheng{
Concretely, we first execute every within-budget candidate on a small data sample. We aggregate their sample rankings into a proxy ground truth using reciprocal rank fusion (RRF)~\cite{RRF}, which scores each item by its rank in every run and orders items by the summed score. We note that there are other alternative aggregation mechanism such as the Borda Count~\cite{Emerson2013Borda}. 
RRF scores each item using the reciprocal of its rank, giving greater weight to agreement at the top of the ranking while reducing the influence of weak candidates that place relevant items far down the list, which linear positional methods such as Borda count may overemphasize.
We then measure each candidate's agreement with this proxy. Finally, we greedily select the highest-agreement candidates whose combined estimated cost fits the budget, execute each on the full input, and fuse their outputs, again via RRF, into the returned ranking; when the budget admits only one candidate, this reduces to selecting the single most consensus-aligned algorithm. By aggregating access paths both to \emph{estimate} quality and to \emph{produce} the final ranking, the optimizer is robust to the errors of any individual access path.}

\section{Experiments}
\label{sec:experiment}
\begin{figure*}[htbp]
    \centering

    \begin{tabular}{@{}c@{\hspace{1em}}l@{}}
        \rotatebox{90}{Llama3.1-70b} &
        \begin{minipage}{0.98\textwidth}
            \begin{subfigure}{0.24\linewidth}
                \centering
                \includegraphics[width=\linewidth]{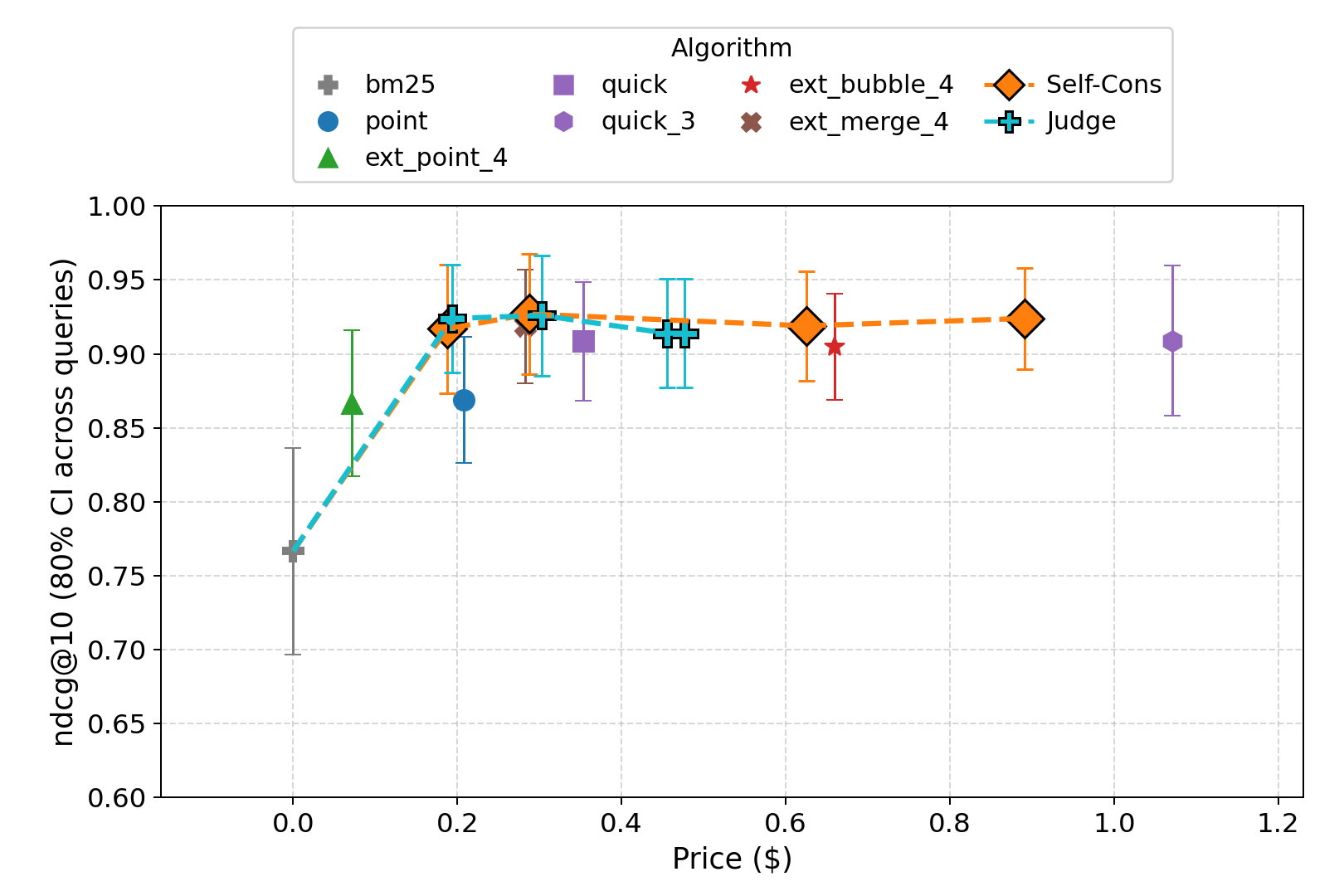}
                \caption{SembenchMovie}
            \end{subfigure}
            \hfill
            \begin{subfigure}{0.24\linewidth} 
                \centering
                \includegraphics[width=\linewidth]{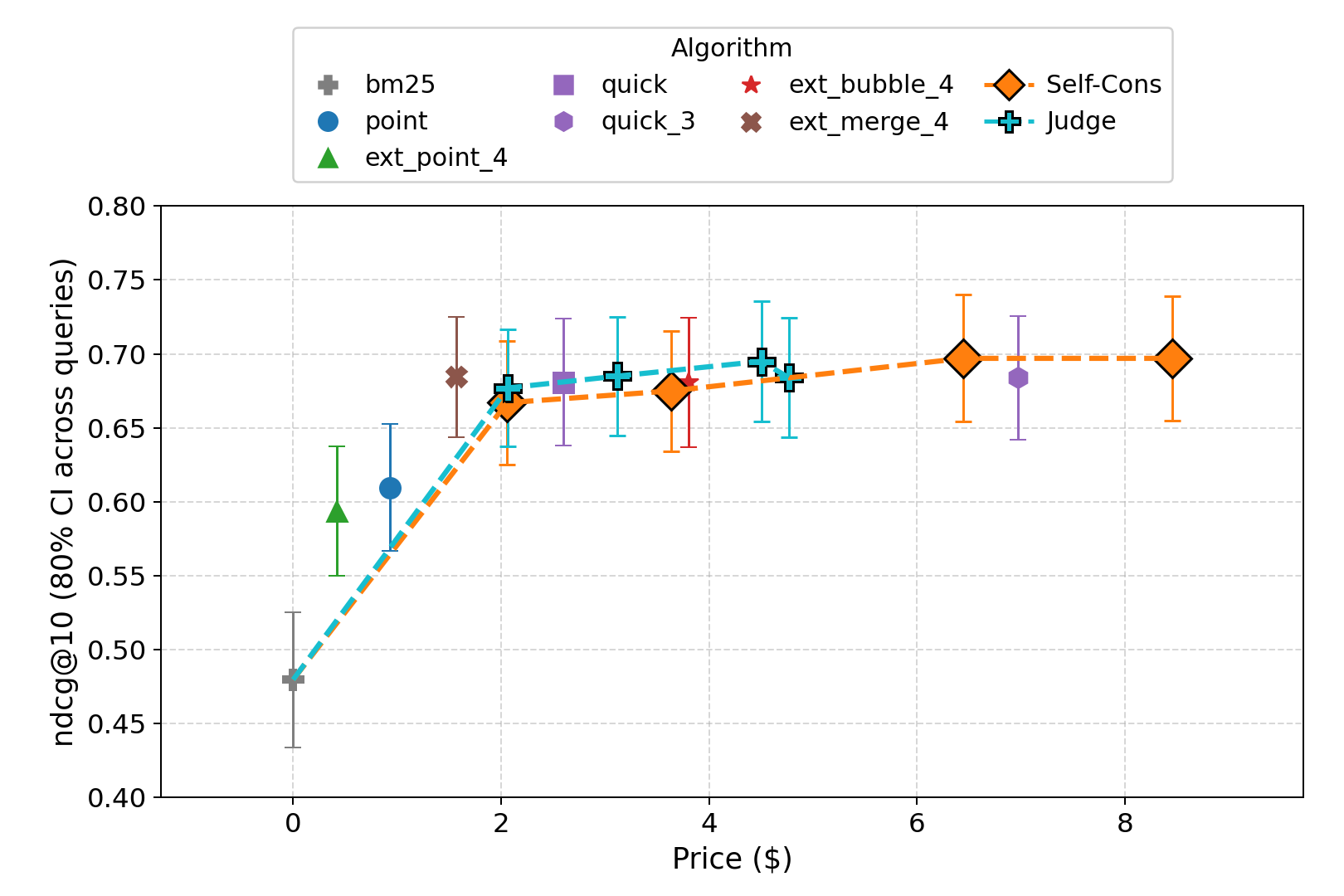}
                \caption{DL20}
            \end{subfigure}
            \hfill
            \begin{subfigure}{0.24\linewidth} 
                \centering
                \includegraphics[width=\linewidth]{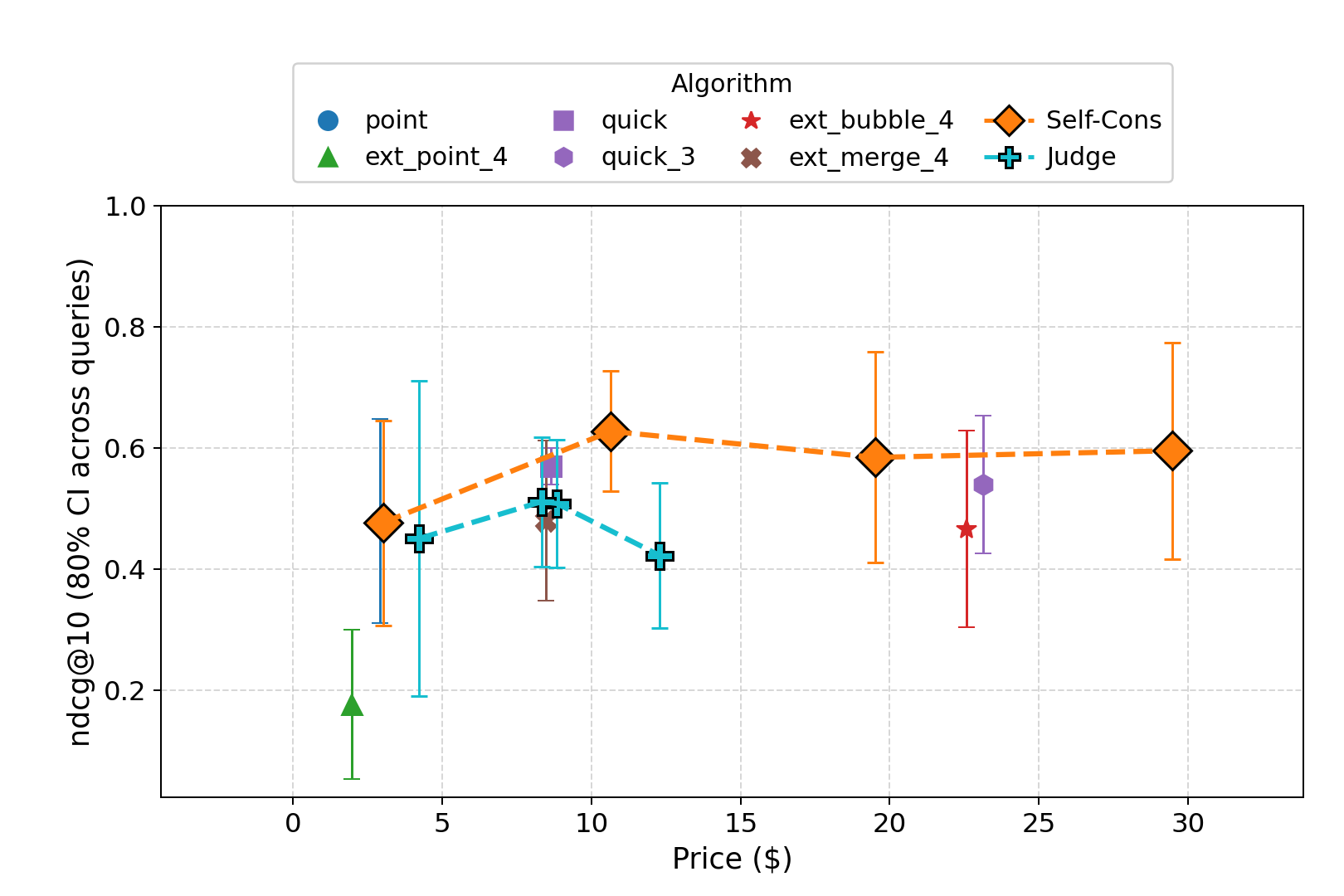}
                \caption{NFCorpus}
            \end{subfigure}
            \hfill
            \begin{subfigure}{0.24\linewidth} 
                \centering
                \includegraphics[width=\linewidth]{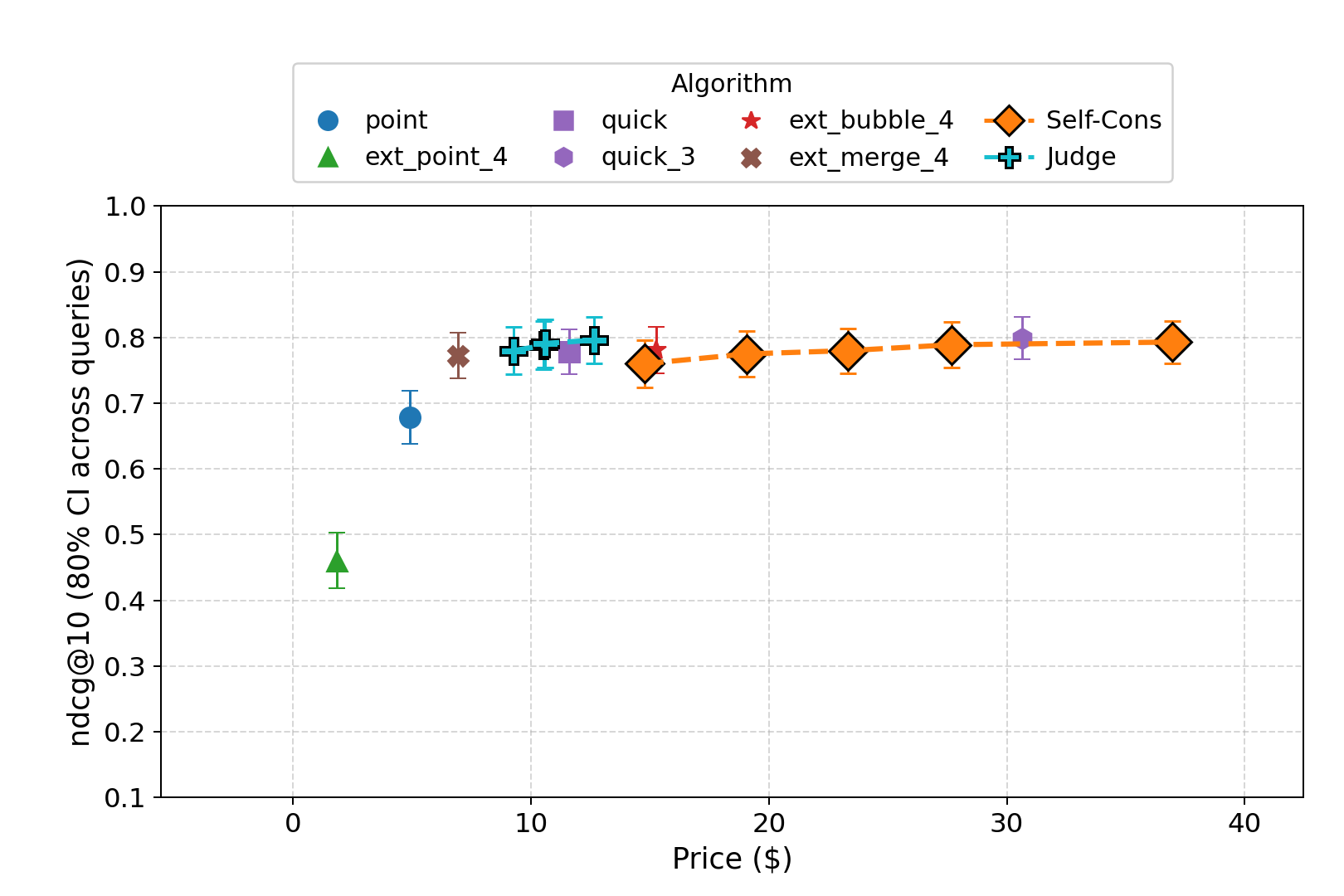}
                \caption{Hellaswag}
            \end{subfigure}
            
        \end{minipage}
    \end{tabular}

    \vspace{1em} 

    \begin{tabular}{@{}c@{\hspace{1em}}l@{}}
        \rotatebox{90}{GPT-5 mini} &
        \begin{minipage}{0.98\textwidth}
            \begin{subfigure}{0.24\linewidth}
                \centering
                \includegraphics[width=\linewidth]{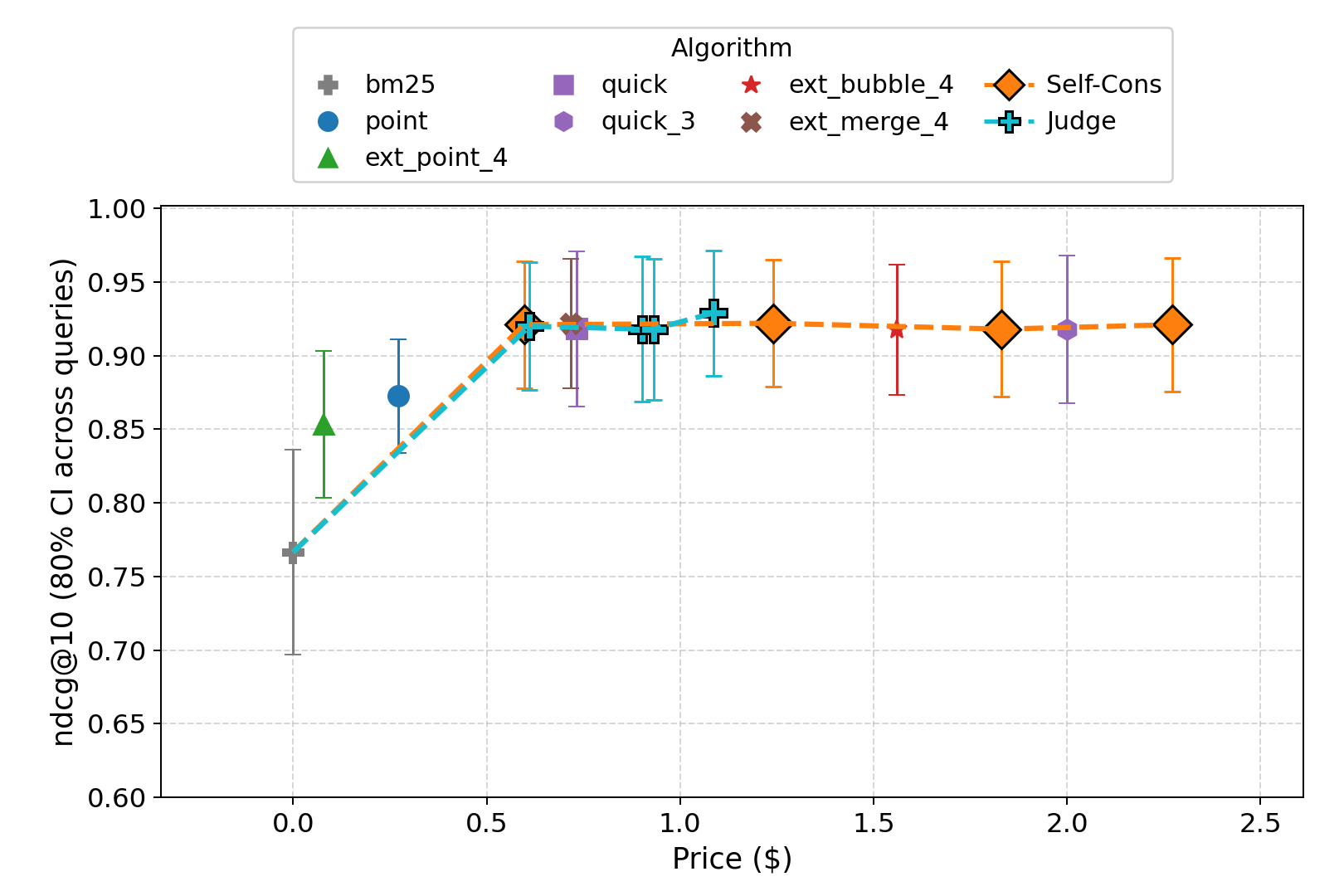}
                \caption{SembenchMovie}
            \end{subfigure}
            \hfill
            \begin{subfigure}{0.24\linewidth} 
                \centering
                \includegraphics[width=\linewidth]{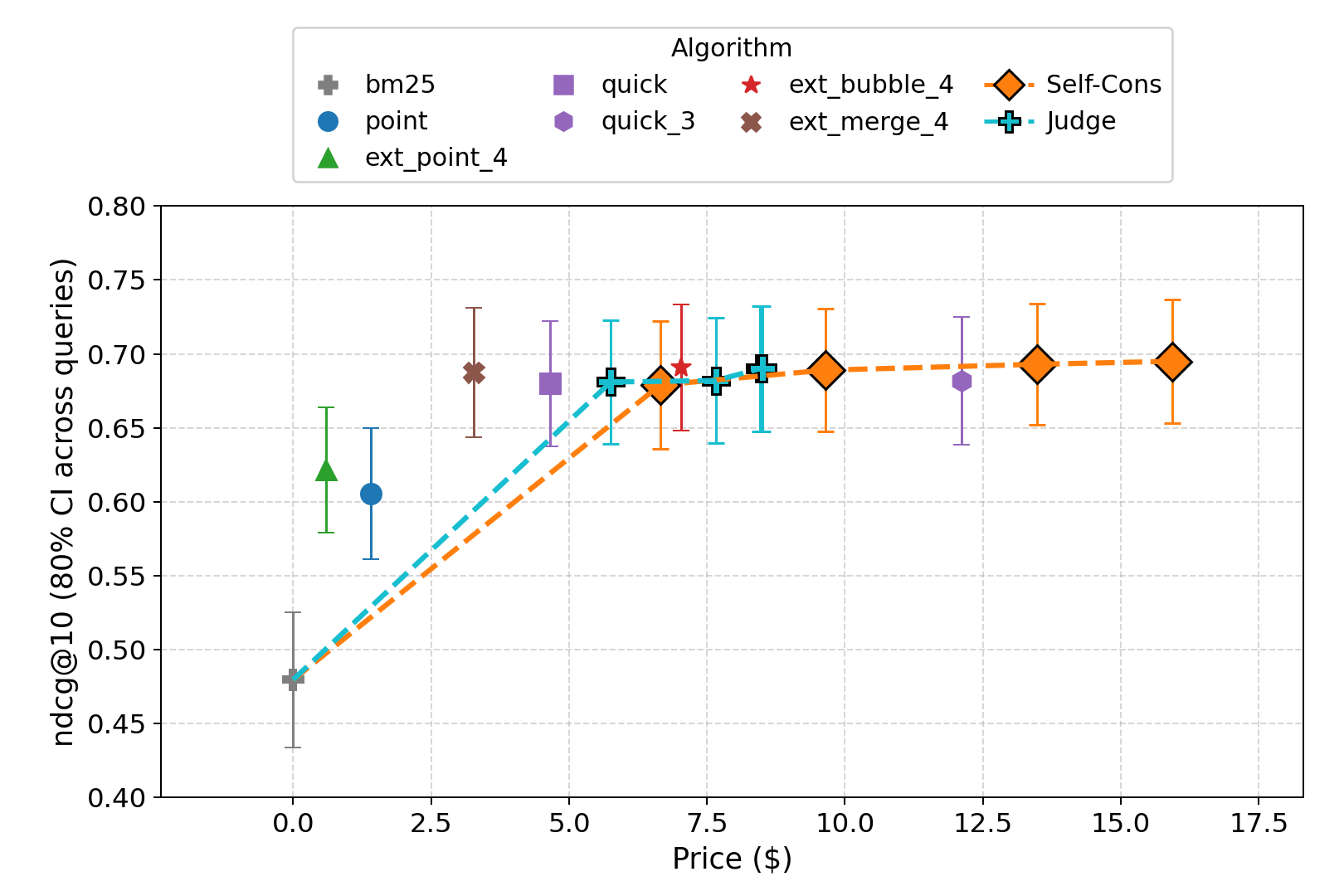}
                \caption{DL20}
            \end{subfigure}
            \hfill
            \begin{subfigure}{0.24\linewidth} 
                \centering
                \includegraphics[width=\linewidth]{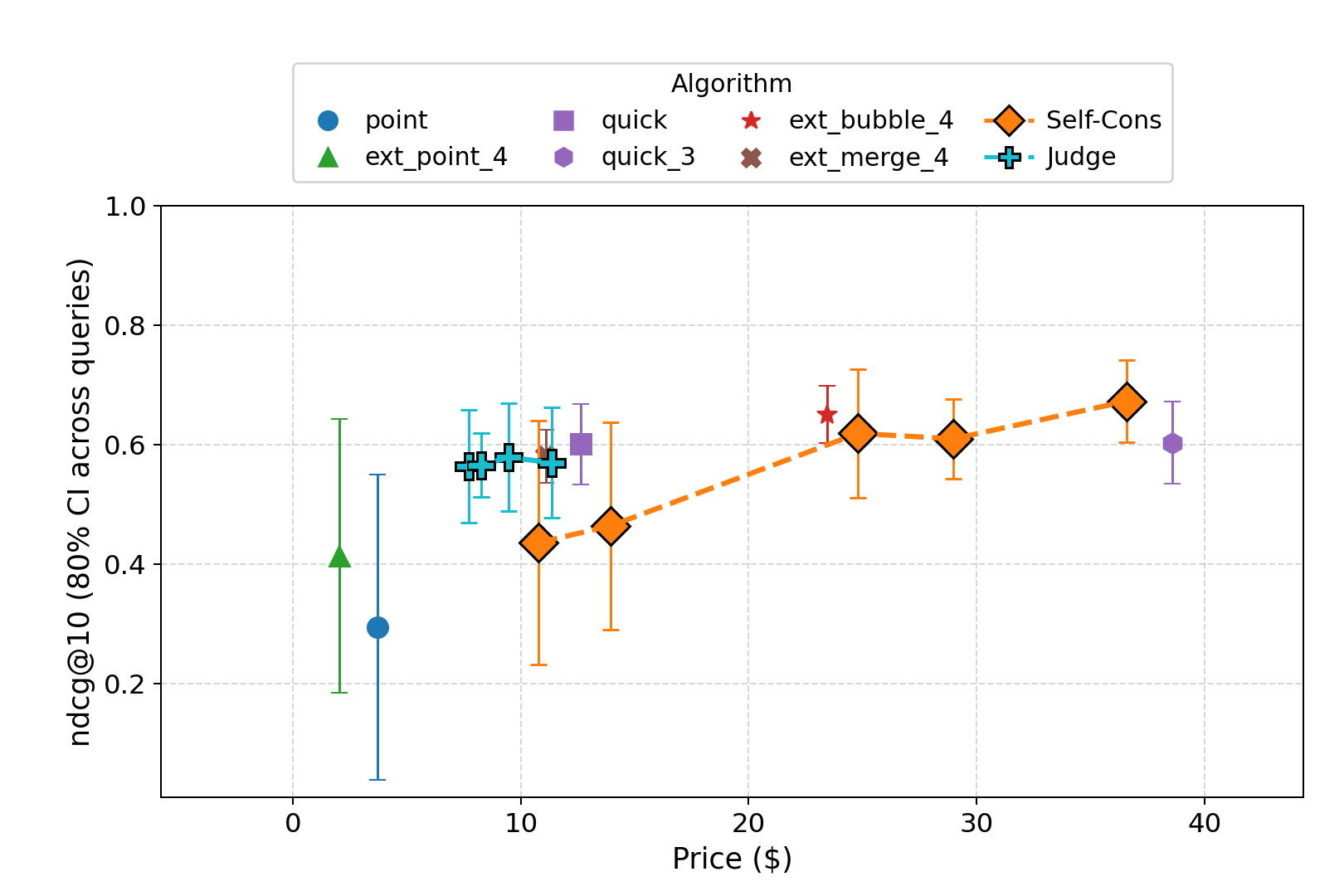}
                \caption{NFCorpus}
            \end{subfigure}
            \hfill
            \begin{subfigure}{0.24\linewidth} 
                \centering
                \includegraphics[width=\linewidth]{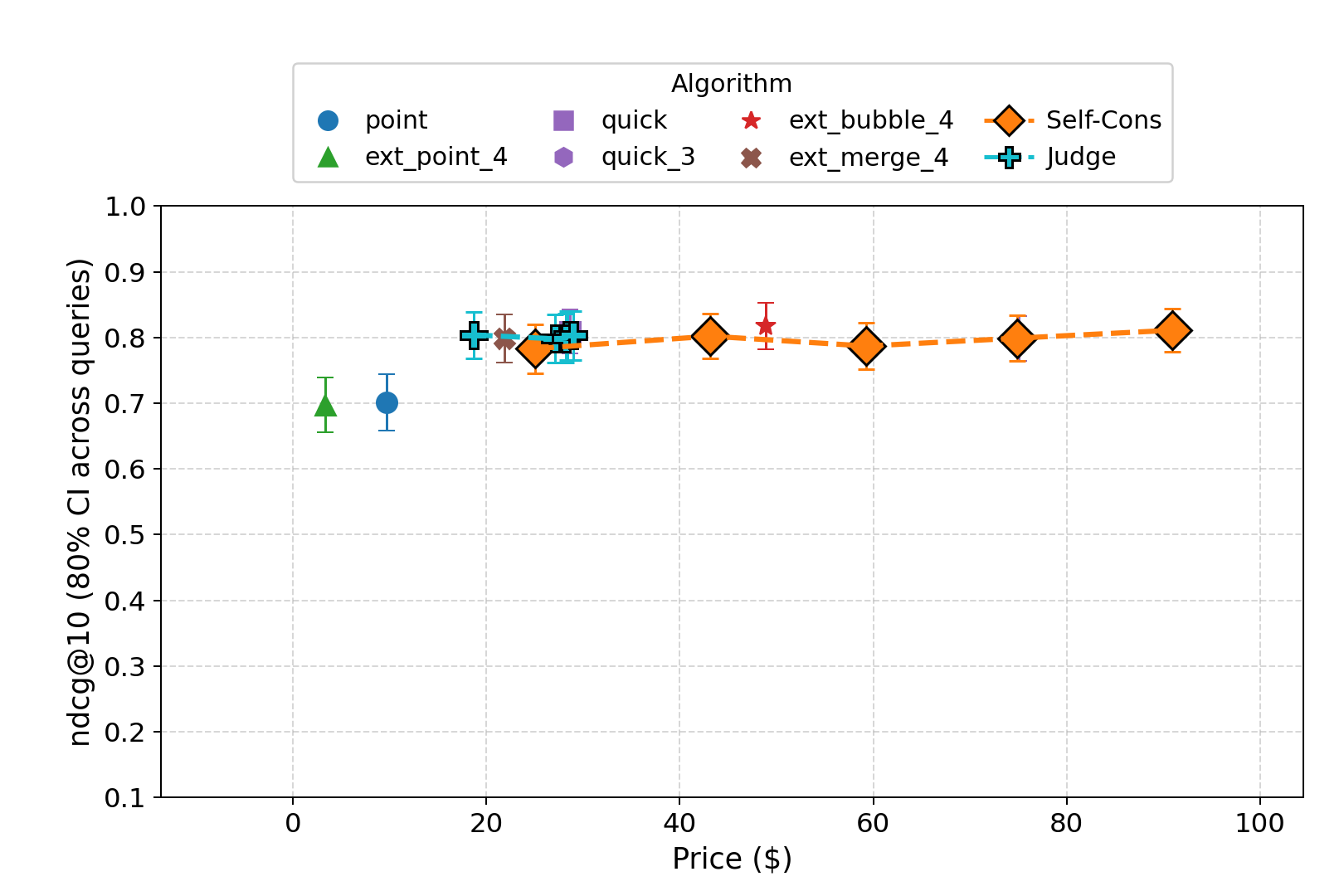}
                \caption{HellaSwag}
            \end{subfigure}
        \end{minipage}
    \end{tabular}

        \vspace{1em} 

    \begin{tabular}{@{}c@{\hspace{1em}}l@{}}
        \rotatebox{90}{Haiku-4.5} &
        \begin{minipage}{0.98\textwidth}
            \begin{subfigure}{0.24\linewidth}
                \centering
                \includegraphics[width=\linewidth]{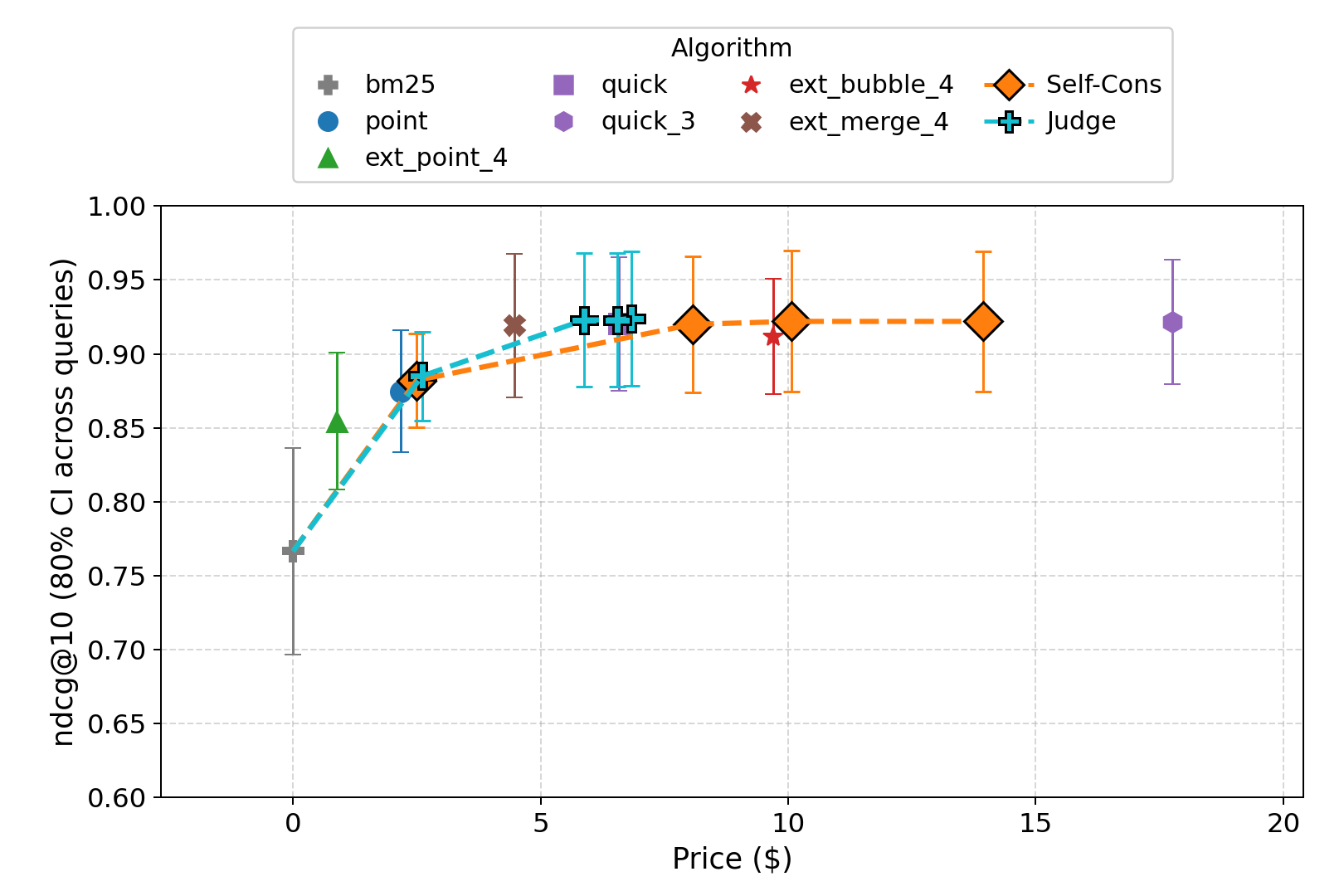}
                \caption{SembenchMovie}
            \end{subfigure}
            \hfill
            \begin{subfigure}{0.24\linewidth} 
                \centering
                \includegraphics[width=\linewidth]{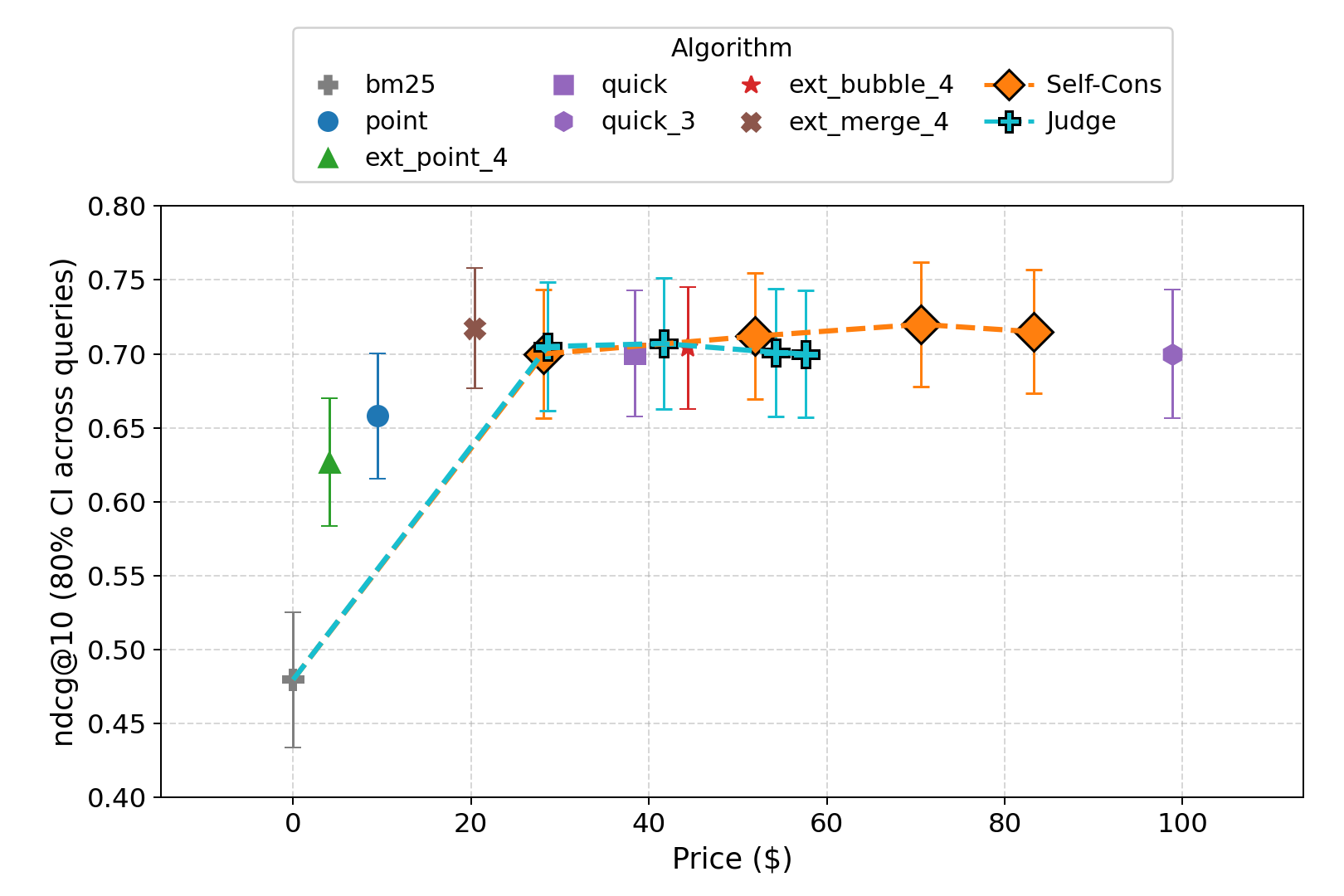}
                \caption{DL20}
            \end{subfigure}
            \hfill
            \begin{subfigure}{0.24\linewidth} 
                \centering
                \includegraphics[width=\linewidth]{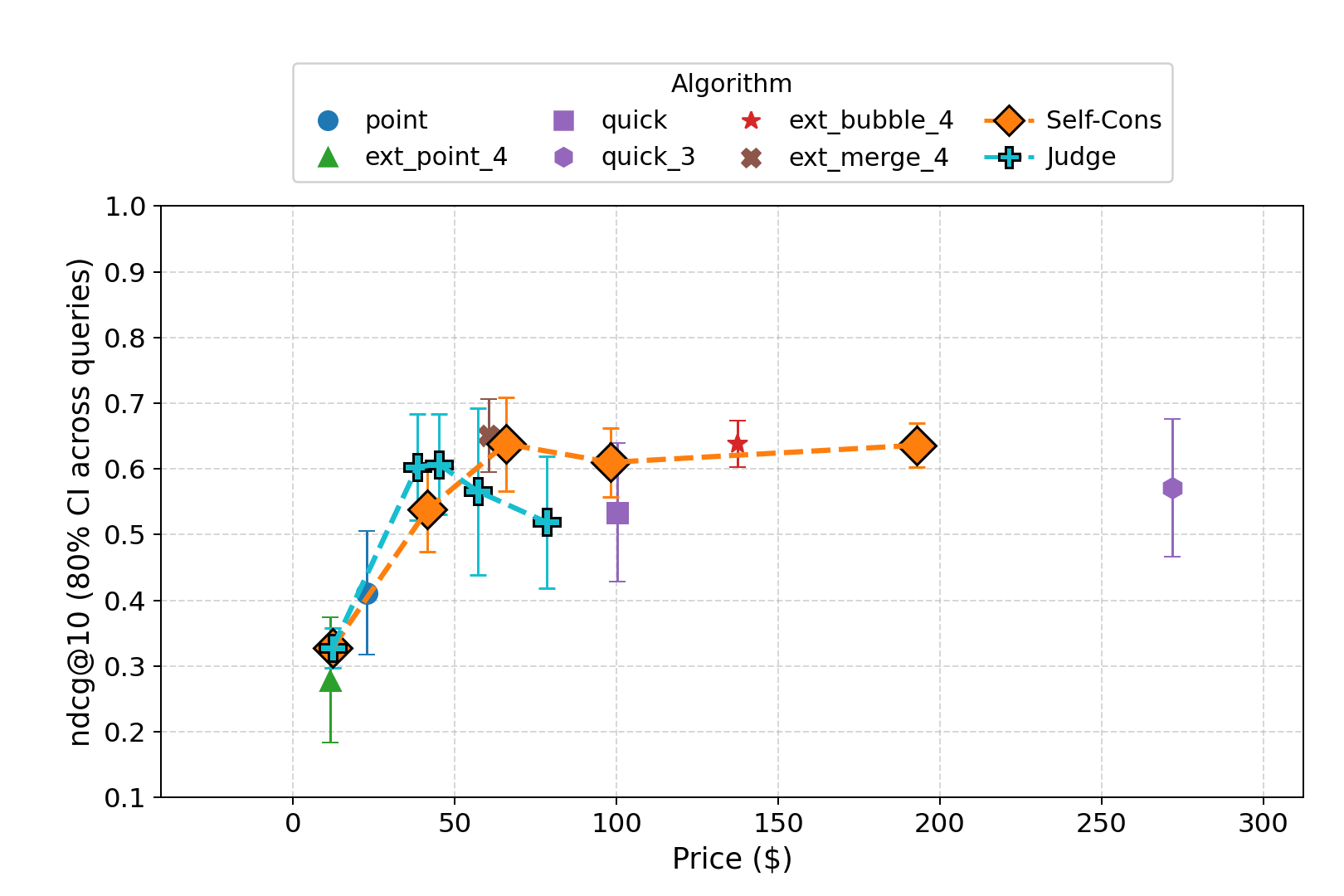}
                \caption{NFCorpus}
            \end{subfigure}
            \hfill
            \begin{subfigure}{0.24\linewidth} 
                \centering
                \includegraphics[width=\linewidth]{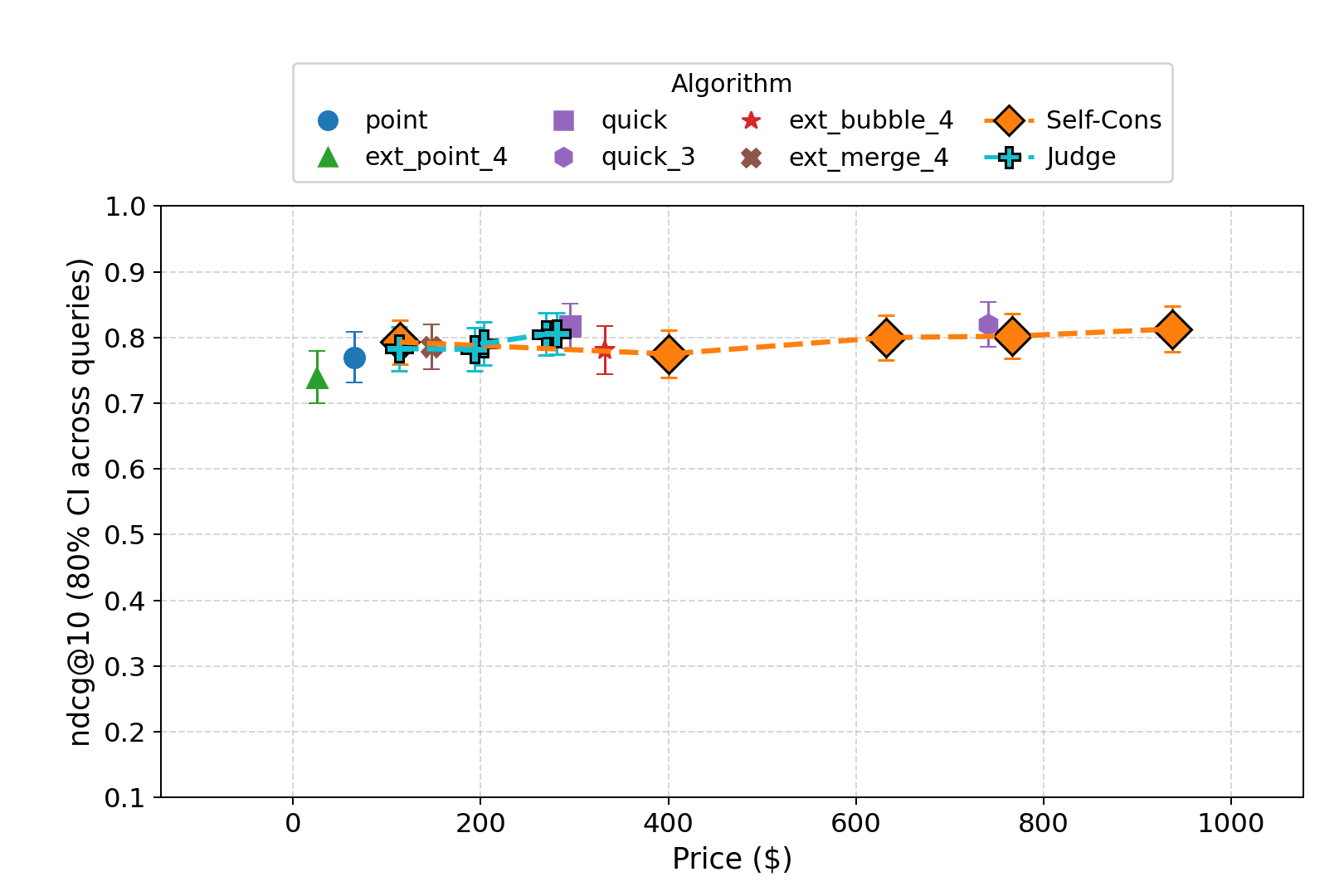}
                \caption{HellaSwag}
            \end{subfigure}
        \end{minipage}
    \end{tabular}

    \caption{Sorting accuracy vs. the monetary budget (\$).}
    \label{fig:main_results}
\end{figure*}

\begin{table*}[t]
\centering
\small
\setlength{\tabcolsep}{4pt}
\begin{tabular}{ll lclclc cc}
\toprule
\multirow{2}{*}{Dataset} & \multirow{2}{*}{Model} & \multicolumn{6}{c}{Top-3 Best Standalone Algorithms} & \multicolumn{2}{c}{ Optimizer Best} \\
\cmidrule(lr){3-8} \cmidrule(l){9-10}
& & 1st Alg & Acc. & 2nd Alg & Acc. & 3rd Alg & Acc. & Judge & Self-Cons. \\
\midrule
\multirow{3}{*}{World}
& Llama-3.1 70B & point\_search & 0.982 & point & 0.975 & ext\_point\_search & 0.969 & \textbf{0.982} & \textbf{0.982} \\
& GPT-5 mini & point\_search & 0.996 & point & 0.996 & quick\_3 & 0.990 & \textbf{0.996} & \textbf{0.996} \\
& Haiku-4.5 & point\_search & 0.98 & ext\_bubble\_4 & 0.98 & quick\_3 & 0.98 & \textbf{0.98} & \textbf{0.98} \\
\midrule

\multirow{3}{*}{Sembench} 
& Llama-3.1 70B & ext\_merge\_4 & 0.919 & quick\_3 & 0.909 & quick & 0.909 & \rev{\textbf{0.926}} & \rev{\textbf{0.927}} \\
& GPT-5 mini & ext\_merge\_4 & 0.922 & quick & 0.918 & quick\_3 & 0.918 & \rev{\textbf{0.929}} & \rev{\textbf{0.922}} \\
& Haiku-4.5 & quick\_3 & 0.922 & quick & 0.920 & ext\_merge\_4 & 0.919 & \rev{\textbf{0.926}} & \rev{\textbf{0.922}} \\

\midrule
\multirow{3}{*}{DL20} 
& Llama-3.1 70B & ext\_merge\_4 & 0.685 & quick\_3 & 0.684 & quick & 0.681 & \rev{\textbf{0.695}} & \rev{\textbf{0.697}} \\
& GPT-5 mini & ext\_bubble\_4 & 0.691 & ext\_merge\_4 & 0.687 & quick\_3 & 0.682 & \rev{0.690} & \rev{\textbf{0.695}} \\
& Haiku-4.5 & ext\_merge\_4 & 0.717 & ext\_bubble\_4 & 0.704 & quick & 0.700 & \rev{0.707} & \rev{\textbf{0.720}} \\
\midrule

\multirow{3}{*}{NfCorpus} 
& Llama-3.1 70B & \rev{quick} & \rev{0.570} & \rev{quick\_3} & \rev{0.540} & \rev{ext\_merge\_4} & \rev{0.480} & \rev{0.512} & \rev{\textbf{0.628}} \\
& GPT-5 mini & ext\_bubble\_4 & \rev{0.651} & quick\_3 & \rev{0.604} & quick & \rev{0.601} & \rev{0.579} & \rev{\textbf{0.673}} \\
& Haiku-4.5 & \rev{ext\_merge\_4} & \rev{0.651} & \rev{ext\_bubble\_4} & \rev{0.638} & quick\_3 & \rev{0.572} & \rev{0.607} & \rev{0.638} \\
\midrule

\multirow{3}{*}{HellaSwag} 
& Llama-3.1 70B & quick\_3 & 0.799 & ext\_bubble\_4 & 0.781 & quick & 0.779 & \rev{0.796} & \rev{0.793} \\
& GPT-5 mini & ext\_bubble\_4 & 0.818 & quick & 0.809 & ext\_merge\_4 & 0.799 & \rev{0.804} & \rev{0.811} \\
& Haiku-4.5 & quick\_3 & 0.820 & quick & 0.817 & ext\_merge\_4 & 0.786 & \rev{0.806} & \rev{0.813} \\

\bottomrule
\end{tabular}
\caption{Top-3 single-algorithm results and best optimizer results on each dataset/model pair.}
\label{tab:main-results}
\end{table*}

We present an evaluation of our proposed LLM-as-Judge and Self-Consistency optimizers (Section~\ref{sec:optimizer}), building on top of value-based and comparison-based LLM sorting algorithms (Section~\ref{sec:sorting}).
With systematic evaluation, we demonstrate that our proposed optimizer can achieve near-optimal accuracy through online evaluation.

\subsection{Experimental Setup}

\noindent \textbf{Evaluation Datasets.}
\begin{itemize}
    \item 2020 World Population~\cite{tanuprabhu2020population}: The query is to order these countries by their population.
    
    \item SembenchMovie Q9~\cite{lao2025sembenchbenchmarksemanticquery}: Originally designed to rank items for \textit{Ant-Man and the Wasp} by positivity, we expanded this task to include 5 distinct queries corresponding to the top 5 most reviewed movies.

    \item DL20~\cite{Craswell2021TRECDL2020}: Contains 54 search queries associated with the 8.8 million passage corpus.

    \item \fuheng{NFCorpus~\cite{boteva2016full}: A full-text medical information-retrieval benchmark. }\rev{Due to the high cost, we use three test queries and each query ranks the full 3.6k document corpus.}
        
    \item \fuheng{HellaSwag~\cite{zellers2019hellaswag}:A commensense sentence-completion benchmark. We use the first 100 queries with a corpus of 400 passages, following the same experimental setting as the Google proxy model~\cite{googleproxy}.}
    
\end{itemize}

\noindent \textbf{Evaluation Metrics.}
\fuheng{We use the kendall's tau metric to measure the accuracy of the sorted list in the 2020 world population, and use the $nDCG@10$ metric for the remaining datasets. For each query in the remaining datasets, we appended a DESC LIMIT 10 clause.} For both metrics, closer to 1.0 indicates better sorting quality.

\noindent \textbf{Standalone Algorithms.}
In this section, we compare our proposed optimizer against a comprehensive set of single-algorithm solutions. For the \textbf{value-based} category, we include both external pointwise and standard pointwise algorithms. The pointwise method is implemented as the default in several academic prototypes and utilized by BigQueryfor LLM-based semantic ranking~\cite{lao2025sembenchbenchmarksemanticquery, zhao2024hybrid, googleproxy}. For instance, we employ the exact pointwise prompt utilized by BigQuery for the SembenchMovie dataset.
For the \textbf{comparison-based} category, we evaluate standard quick sort~\cite{qin2023large} and external bubble sort~\cite{sun2023chatgpt}, alongside with our proposed semantic ranking methods: quick sort with majority voting and external merge sort. We also note that Lotus~\cite{patel2024lotus} utilizes the quick sort method. 
The fundamental distinction between our proposed optimizer and these baselines is that while single-algorithm solutions apply a static strategy regardless of the input, our optimizer dynamically selects the near-optimal algorithm at runtime.

\noindent \textbf{Implementation Details.}
\fuheng{Our approach does not require fully materializing the input to the \textsc{LLM Order By} operator before optimization. Instead, we adopt a strategy similar to Palimpzest~\cite{liu2025palimpzest}: the system first executes the upstream portion of the query on a small sample of input rows and uses the resulting intermediate tuples to estimate the cost and quality statistics required by our optimizer. Moreover, LLM responses produced during this sampling phase can be stored in a response cache, also implemented in many other systems~\cite{zhao2024hybrid,patel2024lotus,liu2025palimpzest,LiskowskiCompositionalOL}.}

\fuheng{For all experiments in this section, we directly accessed LLMs via REST API. The temperature is set to 0 to enable greedy decoding and reasoning effort is set to minimal.} 
The sample size for factual world data inquiry stage is fixed to five, and the sample size for running candidate algorithms is fixed to 20. 

\fuheng{We define the user-specified ranking budget $B$ as the maximum monetary cost the optimizer may spend to rank the input items. The cost of each access path is estimated from its required LLM API calls using the corresponding model's API pricing, and the optimizer considers only paths whose estimated cost is within $B$. We evaluate the optimizer across a range of $B$ values to study the resulting quality--cost tradeoff.} 
\fuheng{In the LLM-as-Judge optimizer, the evaluation phase employs the identical backbone model used for the initial sorting. Utilizing the same model facilitates an internal reflection mechanism: the judge can seamlessly re-evaluate the candidate rankings for subtle errors and iteratively refine its assessments to maximize final accuracy.
For the self-consistency optimizer, we measure candidate quality by comparing outputs to a derived aggregation through RRF~\cite{RRF}. The final output is the RRF fusion of the rankings produced by the top budget-feasible candidates by proxy agreement.}

\subsection{Optimizer Main Results}
Figure~\ref{fig:main_results} and Table~\ref{tab:main-results} present our main evaluation. Figure~\ref{fig:main_results} has monetary cost on the $x$-axis and ranking quality on the $y$-axis (higher is better). Each dot also depicts a 80\% confidence interval.
Table~\ref{tab:main-results} shows the top 3 best standalone algorithms and the best score acheived by our proposed optimizers. We sweep a range of user budgets and evaluate both proposed optimizers: the LLM-as-a-judge optimizer and the self-consistency (RRF) optimizer. \rev{Our optimizer is consistently competitive with the strongest standalone algorithm on every dataset: it matches or exceeds its ranking quality on 10 of the 15 dataset/model pairs and stays within $0.02$ of it on the rest.} In
particular, the self-consistency optimizer, which aggregates several access paths rather than committing to one, most reliably matches or surpasses the best single algorithm.

\noindent \textbf{World Population Benchmark.}
In the world population benchmark, across all three models, our optimizer using the inquiry prompt (Prompt~\ref{lst:inquiry_prompt}) successfully identified that the world population is factual information.
The optimizer automatically routed the query to the web search augmented pointwise method, achieving exceptionally high accuracies of 0.98-0.99 across models. As detailed in Table~\ref{tab:main-results}, while the external pointwise method with web search ranks third and exhibits competitive accuracy, it remains less stable than the standard pointwise approach, which is aligned with our observations in Section~\ref{sec:observation} and prior literature~\cite{cheng2023batchpromptingefficientinference}.

\noindent \textbf{Sembench Movie.}
The Sembench Movie dataset comprises five distinct queries based on most reviewed movies. During the initial inquiry phase, these queries are classified as subjective, non-factual data. Consequently, the optimizer bypassed the default method of pointwise with web serach. 
As illustrated in Figure~\ref{fig:main_results}, relying solely on the pointwise sorting algorithm, as currently adopted in the Sembench baseline~\cite{lao2025sembenchbenchmarksemanticquery}, results in a significant degradation of ranking accuracy. 
\fuheng{For instance, on the Claude Haiku-4.5 model, the standard pointwise method scores 0.875 at cost \$2.18. \rev{In contrast, at a similar price point, the LLM-as-Judge and self-consistency optimizers score 0.885 at \$2.60 and 0.882 at \$2.49, and their best scores are 0.926 and 0.922.} 
Moreover, across all three models, the optimizers correctly gravitate toward the high-accuracy regions dominated by comparison-based algorithms, \rev{matching or outperforming the best single algorithm on all three models} as shown in Table~\ref{tab:main-results}.}


\noindent \textbf{DL20.}
\fuheng{DL20 comprises 54 passage-ranking queries. These queries cannot be resolved from parametric knowledge or web search, and require the model to directly assess (passage, query) relevance. As the second column of Figure~\ref{fig:main_results} shows, value-based approaches (pointwise and external pointwise) consistently underperform: across all three models their accuracy plateaus around $0.60$--$0.66$, whereas comparison-based algorithms reach $0.68$--$0.72$. This confirms our earlier observation that complex relevance judgments demand pairwise/listwise comparison and reasoning, rather than independent per-item scoring. Without any ground-truth labels, both optimizers correctly gravitate toward this high-accuracy comparison-based region: on Claude Haiku-4.5, for instance, \rev{they attain $0.707$--$0.720$, while the best single algorithm (our proposed external merge sort) scores $0.717$}, with a similar pattern on the other two models (Table~\ref{tab:main-results}). 
Comparing the two optimizers, \rev{the self-consistency optimizer exceeds the best standalone algorithm on all three models}, while the LLM-as-judge optimizer stays close behind. This quality edge comes at a cost: because the self-consistency optimizer fuses several access paths via RRF whereas the judge executes only the single path it selects, the former is consistently more expensive---on Claude Haiku-4.5, for example, \rev{it spends \$71 to reach $0.720$, versus \$42 for the LLM-as-judge at $0.707$}.}

\noindent \textbf{NFCorpus.}
NFCorpus is our most challenging benchmark: \rev{each query ranks the entire 3.6k document corpus, of which only a few dozen are relevant, so the ranking signal is sparse and noisy. Absolute accuracy is correspondingly lower than on DL20---the best single algorithm reaches $0.57$--$0.65$ across the three models.} 
This difficulty manifests as high variance across access paths. \rev{On Llama-3.1 70B, standalone accuracy spans a wide range ($0.18$--$0.57$): external pointwise scoring collapses, and even the comparison-based sorts disagree widely (quick sort $0.570$ vs.\ external merge sort $0.480$). Such variance makes label-free selection especially hard.}

\fuheng{In an ideal world, ranking accuracy would increase monotonically with the allocated budget. While standalone algorithms exhibit a broadly positive relationship between computation and accuracy, this trend is not absolute: some higher-cost algorithms are in fact \emph{less} accurate. As the budget grows, such pricier-but-weaker algorithms become affordable and enter the candidate pool, and an optimizer that commits to a single path may select one of them, as shown in the LLM-as-judge optimizer in Figure~\ref{fig:main_results}(k). 
The self-consistency optimizer is more robust: by aggregating several access paths through RRF rather than committing to one, it averages out any single poor choice and yields a smoother, near-monotone curve. \rev{On Claude Haiku-4.5, the judge optimizer peaks early and then degrades as pricier-but-weaker paths become affordable, whereas the self-consistency optimizer rises to $0.64$ and stays there. Llama-3.1 70B shows the same pattern: the judge dips to $0.42$ at the third budget while self-consistency peaks at $0.63$, above the best standalone ($0.57$). On GPT-5 mini the judge plateaus around $0.57$ while self-consistency climbs to $0.673$, surpassing the best standalone algorithm ($0.651$).}
}

\noindent \textbf{HellaSwag.}
\fuheng{HellaSwag is a commonsense sentence completion task: $100$ queries ranks a shared pool of $400$ candidates. The strongest standalone algorithm is also the most expensive: quick sort with
majority voting attains $0.82$ on Claude Haiku-4.5 but costs \$741. \rev{On this dataset the optimizers come close to, but do not exceed, the best single algorithm: the self-consistency optimizer reaches $0.793$, $0.811$ and $0.813$ and the LLM-as-judge $0.796$, $0.804$ and $0.806$ for Llama-3.1 70B, GPT-5 mini and Haiku-4.5, within $0.003$--$0.014$ of the best standalone ($0.799$, $0.818$ and $0.820$). The judge reaches these scores at a fraction of the cost of the best standalone algorithm (e.g., \$281 vs.\ \$741 on Haiku-4.5), since it executes a single cheaper path. HellaSwag has exactly one correct completion per query, so per-query $nDCG@10$ is nearly binary; fusing several rankings through RRF can demote the correct item whenever the fused lists disagree, which limits the gain from self-consistency here, and both optimizers show small non-monotonic dips as the budget grows.}
}

\noindent \textbf{Summary.}
\fuheng{Across all datasets and models, both optimizers recover a near-optimal cost--quality Pareto frontier without ground-truth labels, \rev{matching or closely approaching the best standalone value-based and comparison-based algorithm at a comparable budget.} The two policies occupy complementary regions of this frontier. The LLM-as-judge optimizer relies on internal reflection to sidestep the noise and inconsistent logic of long comparison traces, delivering strong accuracy at low cost; the self-consistency optimizer instead fuses the collective signal of multiple ranking passes to raise the accuracy ceiling, at greater expense.}
Finally, as shown in Table~\ref{tab:main-results}, our proposed external merge sort and quick sort with majority votes are consistently among the strongest standalone algorithms, providing a robust baseline on which the optimization framework builds.

\begin{table*}[t]
\centering
\caption{\rev{Mean absolute \% deviation of the estimated number of LLM calls from the number actually issued by the full run.}}
\label{tab:opt-cost-error}
\begin{tabular}{l|ccc|ccc|ccc}
\toprule
& \multicolumn{3}{c|}{\textbf{DL20}} & \multicolumn{3}{c|}{\textbf{HellaSwag}} & \multicolumn{3}{c}{\textbf{NFCorpus}} \\
\textbf{Alg} & Llama-3.170b & GPT-5mini & Haiku-4.5 & Llama-3.170b & GPT-5mini & Haiku-4.5 & Llama-3.170b & GPT-5mini  & Haiku-4.5 \\
\midrule
ext\_point\_4  & \rev{0.0\%} & \rev{0.0\%} & \rev{0.0\%} & \rev{0.0\%} & \rev{0.0\%} & \rev{0.0\%} & \rev{0.0\%} & \rev{0.0\%} & \rev{0.0\%} \\
point        & \rev{0.0\%} & \rev{0.0\%} & \rev{0.0\%} & \rev{0.0\%} & \rev{0.0\%} & \rev{0.0\%} & \rev{0.0\%} & \rev{0.0\%} & \rev{0.0\%} \\
ext\_merge\_4  & \rev{7.0\%} & \rev{6.4\%} & \rev{6.3\%} & \rev{3.9\%} & \rev{4.0\%} & \rev{3.4\%} & \rev{3.1\%} & \rev{3.6\%} & \rev{3.8\%} \\
ext\_bubble\_4 & \rev{4.2\%} & \rev{4.2\%} & \rev{4.2\%} & \rev{1.0\%} & \rev{1.0\%} & \rev{1.0\%} & \rev{0.2\%} & \rev{0.2\%} & \rev{0.2\%} \\
quick        & \rev{32.0\%} & \rev{33.0\%} & \rev{31.2\%} & \rev{23.4\%} & \rev{29.8\%} & \rev{32.4\%} & \rev{1.7\%} & \rev{18.7\%} & \rev{25.1\%} \\
quick\_3     & \rev{35.0\%} & \rev{37.1\%} & \rev{36.6\%} & \rev{27.6\%} & \rev{34.2\%} & \rev{29.9\%} & \rev{17.2\%} & \rev{18.5\%} & \rev{20.9\%} \\
\bottomrule
\end{tabular}
\end{table*}

\subsection{Latency and Estimated Cost}
\fuheng{In this subsection, we discuss the latency associated with each algorithm, and estimated costs based on Table~\ref{tab:complexity} and samples.}

\begin{figure}
    \centering
    \includegraphics[width=0.8\linewidth]{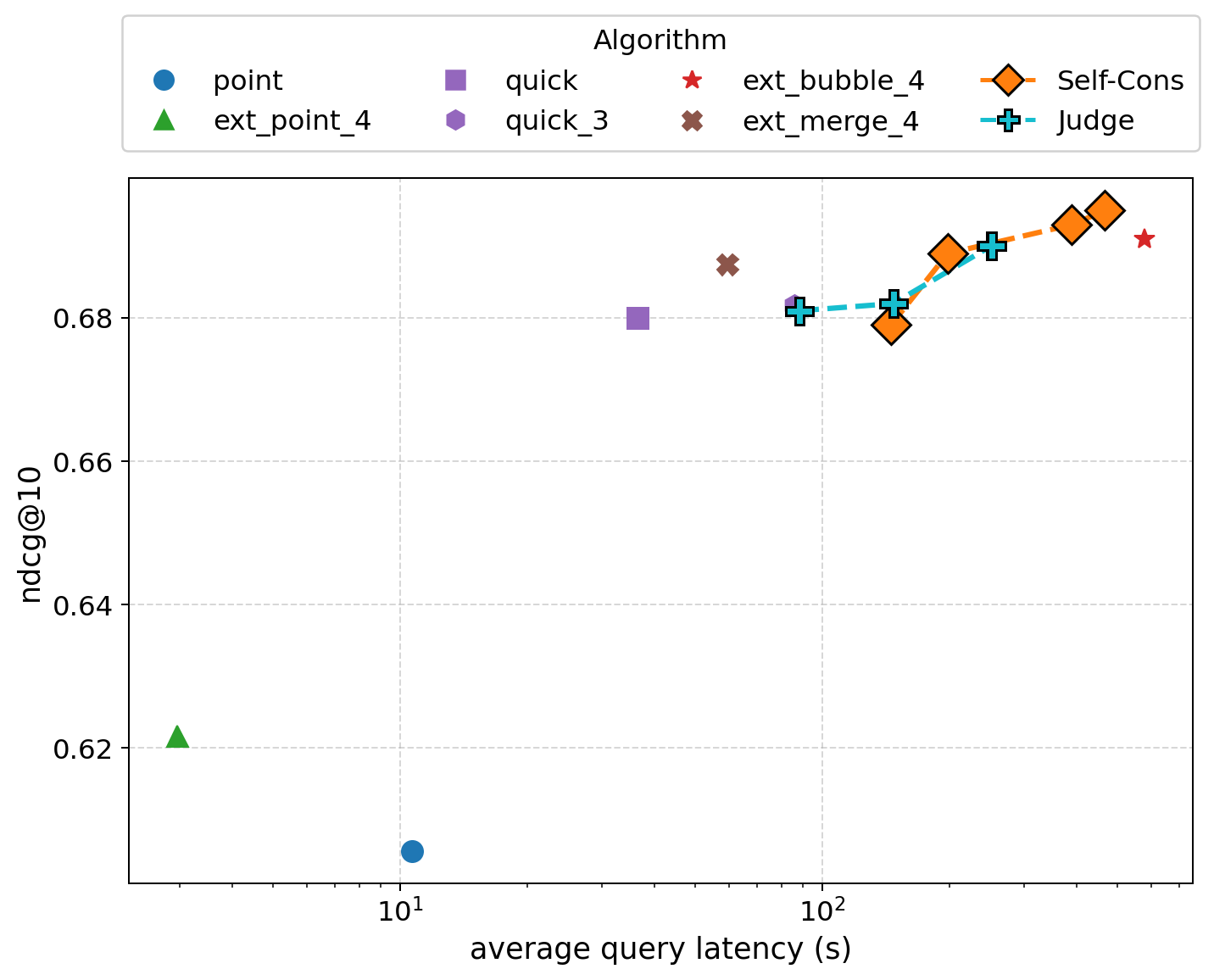}
    \caption{\rev{DL20 GPT-5 mini Latency.}}
    \label{fig:dl20haikulatency}
\end{figure}

\noindent \textbf{Latency.}
\fuheng{Figure~\ref{fig:dl20haikulatency} shows the average latency per query ($x$-axis in log scale) and the quality ($y$-axis), \rev{running GPT-5 mini on DL20}. 
To measure latency, we assume that at most $100$ LLM requests are in flight at any time. We note that service providers often impose rate limits; for example, Snowflake caps Llama-3.1 70b at $200$ requests per minute, whereas Claude Haiku-4.5 has a higher threshold. 
\rev{Value-based methods are the fastest but also the lowest quality. Comparison-based algorithms trade latency for quality: quick sort (37 seconds) is the fastest because the comparisons against each pivot run concurrently, and external merge sort (59 seconds) reaches $0.687$. External bubble sort is a cautionary case. It attains the best standalone quality ($0.691$) but takes $580$s, roughly $10\times$ of our proposed external merge sort, because its passes are inherently sequential. The two optimizers occupy complementary regions of the frontier. The LLM-as-judge reaches $0.690$ at $250$s, since it executes only the single path it selects. The self-consistency optimizer improves monotonically with the budget to the highest quality ($0.695$) at $470$s, higher accuracy and faster than external bubble sort.}}


\noindent \textbf{LLM Calls Estimation.}
\fuheng{The optimizer estimates each candidate's full cost by executing it on $20$ samples and then extrapolate according to the algorithm's LLM-call complexity (Table~\ref{tab:complexity}). \rev{Table~\ref{tab:opt-cost-error} reports, for each algorithm and model on DL20, HellaSwag and NFCorpus, the mean absolute percentage error (MAPE) between the number of LLM calls predicted by the complexity formula and the number actually issued by the full run. The call counts of the value-based paths are exact, and those of external bubble sort and external merge sort are within 0.2--7\%: their formulas only round the window and chunk counts, so the error shrinks as the input grows (for bubble sort from 4.2\% on 100 passages to 0.2\% on 3.6k documents). Quicksort is the main exception, with errors of 2--37\%. Its number of comparisons depends on the pivot choices; quicksort with majority voting exhibits a similar relative error.}
}

\subsection{Ablation Studies}

\begin{figure}
    \centering
    \includegraphics[width=0.8\linewidth]{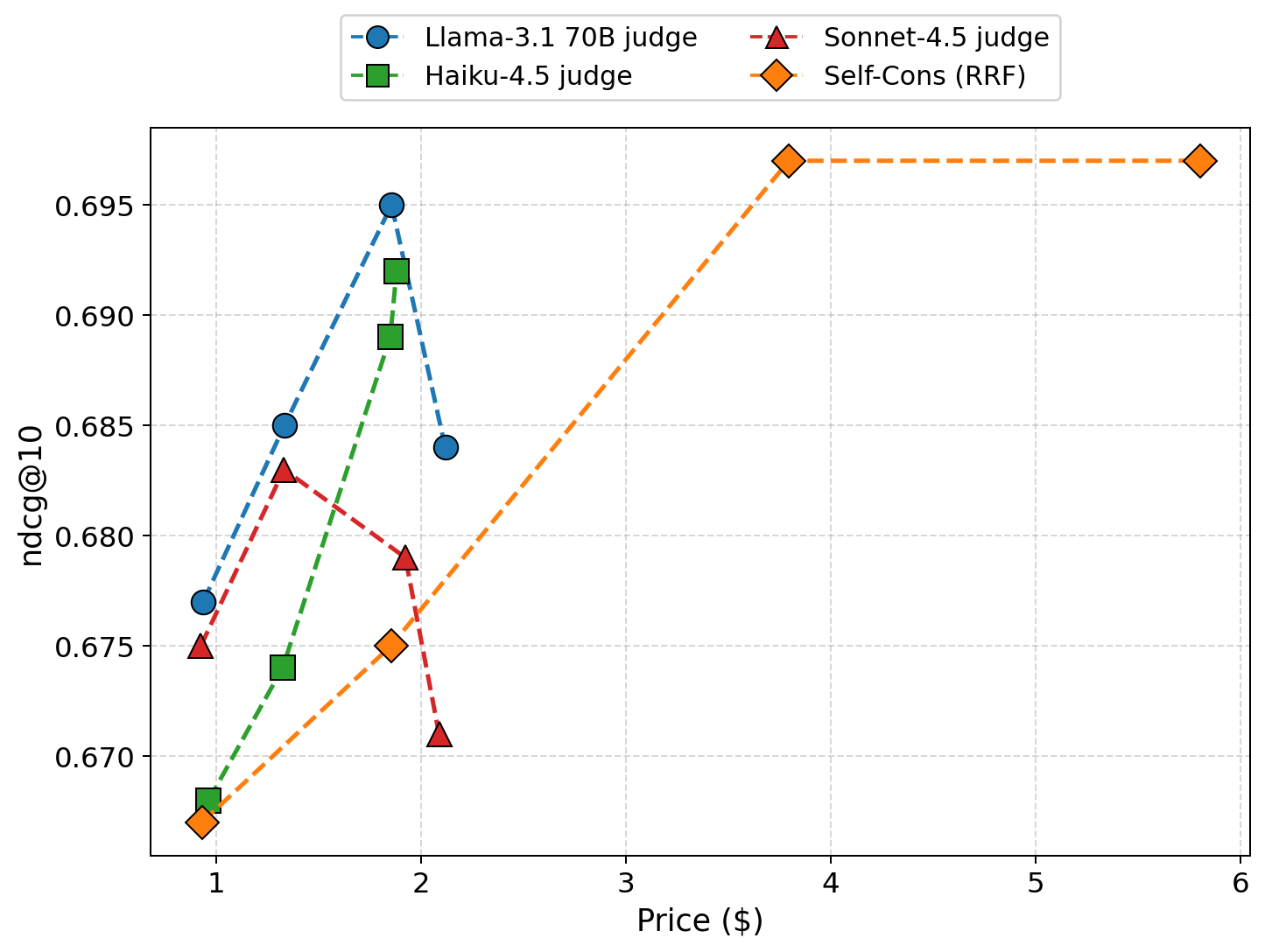}
    \caption{Judge-model ablation on DL20, using Llama-3.1 70b as the ranker.}
\label{fig:judge-ablation-llama}
    \label{fig:judge-ablation-llama}
\end{figure}

\begin{table}[hbtp]
\centering
\caption{Percentage of DL20 queries for which the selected ranking strategy matches or outperforms the best static access path, using Llama-3.1 70b as the ranker.}
\label{tab:judge-vs-static}
\begin{tabular}{lrrrr}
\toprule
Ranking budget & Llama & Haiku & Sonnet & RRF \\
\midrule
\$1 & \rev{16.7\%} & 18.5\% & 18.5\% & \rev{14.8\%} \\
\$2 & \rev{16.7\%} & \rev{20.4\%} & \rev{22.2\%} & \rev{24.1\%} \\
\$4 & \rev{22.2\%} & 25.9\% & \rev{24.1\%} & \rev{31.5\%} \\
\$6 & \rev{22.2\%} & \rev{33.3\%} & \rev{25.9\%} & 33.3\% \\
\bottomrule
\end{tabular}
\end{table}

\noindent \textbf{Judge model.}
\fuheng{Figure~\ref{fig:judge-ablation-llama} isolates the effect of the judge model. We fix Llama-3.1 70B as the ranking model and vary the LLM-as-judge among Llama-3.1 70B, Claude Haiku-4.5, and Claude Sonnet-4.5, with the self-consistency (RRF) optimizer included as a reference. Across ranking budgets, the three judges perform similarly, and using a stronger judge does not consistently improve path selection. \rev{At the largest budget, Haiku-4.5 performs best among the judges ($0.689$), ahead of the larger and more expensive Sonnet-4.5 ($0.671$), while the Llama-3.1 70B self-judge peaks earlier ($0.695$ at the \$4 budget).} This suggests that judge effectiveness is not monotonic in model capability and can vary substantially across models. The self-consistency optimizer is more robust and achieves the highest overall quality (\rev{$0.697$}), albeit at a larger budget because it executes and fuses multiple complete rankings rather than selecting a single access path.}

\fuheng{Table~\ref{tab:judge-vs-static} provides a complementary view by reporting the fraction of queries for which the selected ranking matches or outperforms the best static access path. For all policies, the fraction generally increases with the available budget. \rev{At the largest budget, self-consistency and the Haiku-4.5 judge are the most reliable, matching or outperforming the best static path on $33\%$ of queries, compared with $22\%$--$26\%$ for the other two judge-based policies. Among the judges, the trend is consistent with Figure~\ref{fig:judge-ablation-llama}: Haiku-4.5 performs best ($33.3\%$), ahead of the larger and more expensive Sonnet-4.5 ($25.9\%$) and of the Llama-3.1 70B self-judge ($22.2\%$).} 
}

\begin{figure}[t]
\centering
\includegraphics[width=0.7\linewidth]{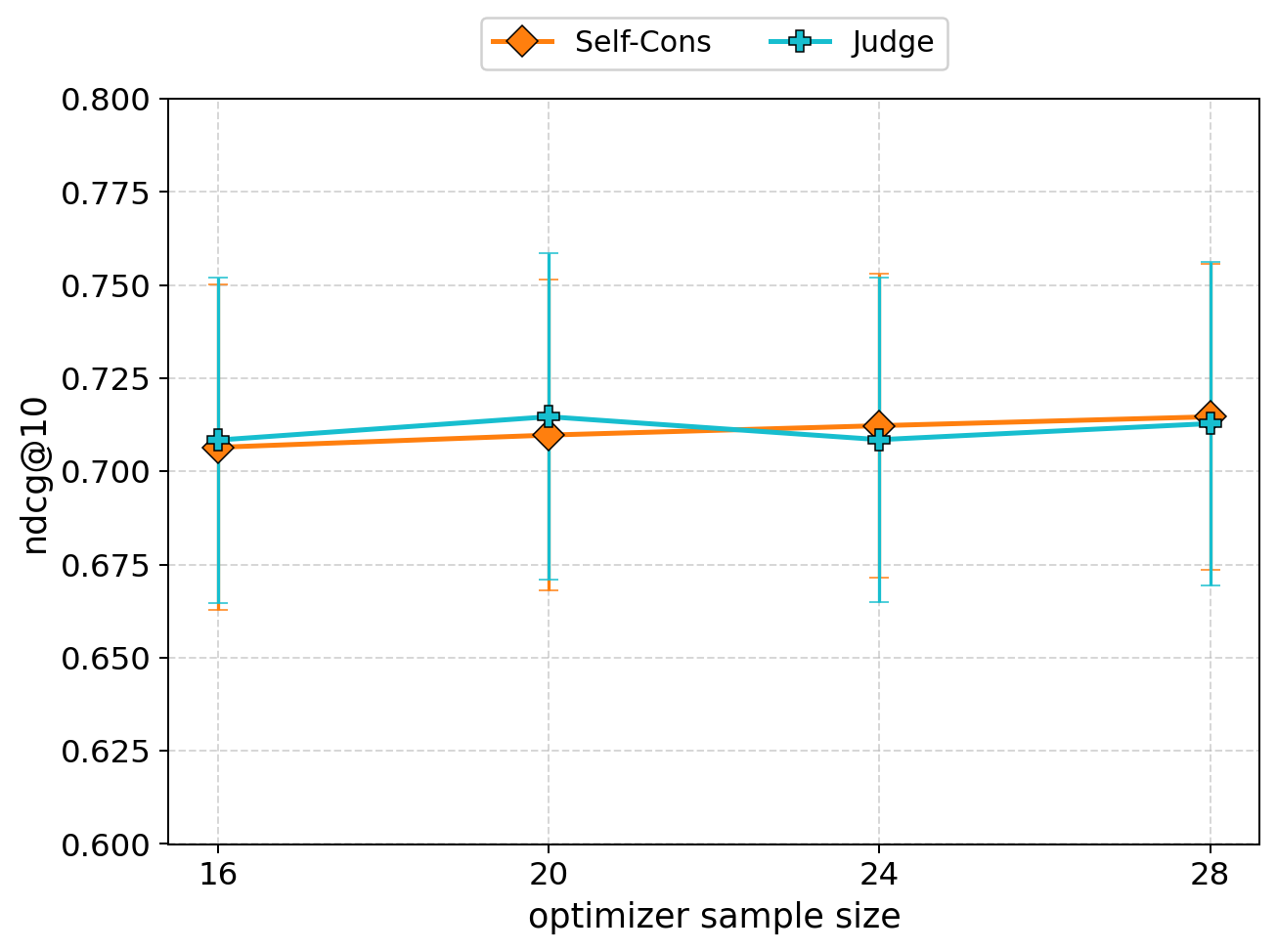}
\caption{Optimizer sample-size ablation on DL20 (base ranker Haiku-4.5).}
\label{fig:samplesize}
\end{figure}

\noindent \textbf{Sample size.}
\fuheng{Figure~\ref{fig:samplesize} varies the number of items the optimizer probes to estimate each access path's quality and cost, holding the base ranker (Haiku-4.5)and budget (\$28) fixed, for both the Self-Consistency and LLM-as-a-Judge selection policies. Increasing the sample size from $16$ to $28$ does not produce a statistically distinguishable change in final NDCG@10: the $80\%$ confidence intervals overlap substantially for both the self-consistency and judge policies. In contrast, the probing costs grow roughly linearly. Since ranking quality does not measurably improve with larger samples, a small
sample of $20$ suffices and is our default.}

\begin{table}[t]
\centering
\caption{Accuracy of the inquiry prompt at distinguishing world knowledge from context-dependent questions}
\label{tab:inquiry}
\begin{tabular}{lc}
\toprule
Model & Accuracy \\
\midrule
Llama-3.1-70B & $88.7\%$ \\
GPT-5-mini    & $90.1\%$ \\
Haiku-4.5     & $88.7\%$ \\
\bottomrule
\end{tabular}
\end{table}

\noindent \textbf{Inquiry Prompt.}
\fuheng{We evaluate how reliably the inquiry prompt separates factual world knowledge (\emph{parametric}) from context-dependent (\emph{non-parametric}) knowledge.
To do so, we construct a balanced benchmark of $274$ questions with ground-truth labels: $137$ factal world knowledge questions drawn from Natural Questions~\cite{kwiatkowski2019natural}. This dataset only contains factual world knowledge questions (e.g., \textit{who wrote the novel Pride and Prejudice''}). Hence, we inlcude another $137$ \emph{non-parametric} questions drawn from SituatedQA~\cite{zhang-choi-2021-situatedqa}, which contain questions that require context information (e.g., \textit{when did we last win a national championship?}).
As shown in Table~\ref{tab:inquiry}, the direct inquiry prompt distinguishes the two knowledge types with $88.7\%$--$90.1\%$ accuracy across three model families. These results suggest that the inquiry prompt provides a robust and model-agnostic mechanism. We also find that the direct inquiry prompt works perfectly on the NBA and world population test datasets.
}

\section{Discussion and Future Work}
\label{sec:discussion}
In this work, we introduced the LLM ORDER BY semantic operator and systematically studied its physical implementations. In addition to adapting existing semantic sorting methods, we developed two new access paths: external merge sort and quicksort with majority voting, which further expand the accuracy-cost design space. External merge sort achieves strong ranking quality at relatively low cost, while quicksort with majority voting can match or exceed its quality only with a substantially larger computational budget. Moreover, our evaluation reveals a consistent pattern: no single access path dominates across all queries and datasets. The best implementation depends on both the task and the underlying data, motivating adaptive access-path selection rather than a fixed execution strategy. We find that sorting quality improves rapidly as the ranking budget increases from low to moderate levels, but the gains diminish and plateau at higher budgets. Building on this observation, we developed an optimizer that selects among access paths under a user-specified ranking budget without requiring ground-truth labels. By probing only a small sample of the input, the optimizer can identify high-quality execution strategies.

Through extensive experiments, we find that our proposed LLM Order By optimizer can select effective access paths without ground-truth labels, recovering near-optimal points on the cost-quality frontier and matching or exceeding the best standalone access path in most settings. 
Its two policies offer complementary trade-offs along this frontier. The self-consistency policy combines multiple access paths to achieve the highest quality, while the LLM-as-judge policy selects a single path and achieves competitive quality at substantially lower cost. 
Overall, these results demonstrate that adaptive, budget-aware access-path selection is a practical execution strategy for the LLM ORDER BY operator.

\fuheng{In future work, we plan to extend the optimizer to explicitly account for ranking latency. While industrial systems often use embedding-based retrieval to first narrow the candidate set to hundreds of relevant documents~\cite{liu2025leveraging, sun2023chatgpt, ma2023zero}, exhaustively ranking millions of documents with LLMs remains prohibitively expensive and slow. A promising direction is to incorporate ideas from Google's proxy-model-based semantic ranking approach~\cite{googleproxy}, which trains a lightweight ranking model online to approximate LLM-based ranking. The lightweight ranking model is orders of magnitude cheaper and faster, at the cost of reduced ranking accuracy relative to the LLM.
Our optimizer could then jointly reason about LLM and proxy-model execution, trading off ranking quality, monetary cost, and latency.
}
Furthermore, we plan to expand the capability of our optimizer to support other operators, such as semantic \texttt{JOIN}~\cite{trummer2025implementingsemanticjoinoperators, sun2025quest}, \texttt{Extract}~\cite{lin2025twixautomaticallyreconstructingstructured, wang2025cequel}, and \texttt{Filter}~\cite{snowflake_cortex_aisql}.

\balance
\bibliographystyle{ACM-Reference-Format}
\bibliography{reference}

@article{zhao2024sphinteract,
  title={Sphinteract: Resolving Ambiguities in NL2SQL through User Interaction},
  author={Zhao, Fuheng and Deep, Shaleen and Psallidas, Fotis and Floratou, Avrilia and Agrawal, Divyakant and Abbadi, Amr El},
  journal={Proceedings of the VLDB Endowment},
  volume={18},
  number={4},
  pages={1145--1158},
  year={2024},
  publisher={VLDB Endowment}
}

@inproceedings{floratou2024nl2sql,
  title={Nl2sql is a solved problem... not!},
  author={Floratou, Avrilia and Psallidas, Fotis and Zhao, Fuheng and Deep, Shaleen and Hagleither, Gunther and Tan, Wangda and Cahoon, Joyce and Alotaibi, Rana and Henkel, Jordan and Singla, Abhik and others},
  booktitle={CIDR},
  year={2024}
}

@article{li2023can,
  title={Can llm already serve as a database interface? a big bench for large-scale database grounded text-to-sqls},
  author={Li, Jinyang and Hui, Binyuan and Qu, Ge and Yang, Jiaxi and Li, Binhua and Li, Bowen and Wang, Bailin and Qin, Bowen and Geng, Ruiying and Huo, Nan and others},
  journal={Advances in Neural Information Processing Systems},
  volume={36},
  pages={42330--42357},
  year={2023}
}

@inproceedings{zhuge2024gptswarm,
  title={Gptswarm: Language agents as optimizable graphs},
  author={Zhuge, Mingchen and Wang, Wenyi and Kirsch, Louis and Faccio, Francesco and Khizbullin, Dmitrii and Schmidhuber, J{\"u}rgen},
  booktitle={Forty-first International Conference on Machine Learning},
  year={2024}
}

@article{zeng2025adl,
  title={ADL: A Declarative Language for Agent-Based Chatbots},
  author={Zeng, Sirui and Yan, Xifeng},
  journal={arXiv preprint arXiv:2504.14787},
  year={2025}
}

@article{zhao2024hybrid,
  title={Hybrid querying over relational databases and large language models},
  author={Zhao, Fuheng and Agrawal, Divyakant and Abbadi, Amr El},
  journal={arXiv preprint arXiv:2408.00884},
  year={2024}
}

@article{shankar2024docetl,
author = {Shankar, Shreya and Chambers, Tristan and Shah, Tarak and Parameswaran, Aditya G. and Wu, Eugene},
title = {DocETL: Agentic Query Rewriting and Evaluation for Complex Document Processing},
year = {2025},
issue_date = {May 2025},
publisher = {VLDB Endowment},
volume = {18},
number = {9},
issn = {2150-8097},
url = {https://doi.org/10.14778/3746405.3746426},
doi = {10.14778/3746405.3746426},
journal = {Proc. VLDB Endow.},
month = may,
pages = {3035–3048},
numpages = {14}
}

@inproceedings{liu2025palimpzest,
  title={Palimpzest: Optimizing ai-powered analytics with declarative query processing},
  author={Liu, Chunwei and Russo, Matthew and Cafarella, Michael and Cao, Lei and Chen, Peter Baile and Chen, Zui and Franklin, Michael and Kraska, Tim and Madden, Samuel and Shahout, Rana and others},
  booktitle={Proceedings of the Conference on Innovative Database Research (CIDR)},
  pages={2},
  year={2025}
}

@inproceedings{he2025cognify,
  title={Cognify: Supercharging gen-ai workflows with hierarchical autotuning},
  author={He, Zijian and Abhyankar, Reyna and Srivatsa, Vikranth and Zhang, Yiying},
  booktitle={Proceedings of the 31st ACM SIGKDD Conference on Knowledge Discovery and Data Mining V. 2},
  pages={932--943},
  year={2025}
}

@article{satriani2025logical,
  title={Logical and Physical Optimizations for SQL Query Execution over Large Language Models},
  author={Satriani, Dario and Veltri, Enzo and Santoro, Donatello and Rosato, Sara and Varriale, Simone and Papotti, Paolo},
  journal={Proceedings of the ACM on Management of Data},
  volume={3},
  number={3},
  pages={1--28},
  year={2025},
  publisher={ACM New York, NY, USA}
}

@article{glenn2024blendsql,
  title={Blendsql: A scalable dialect for unifying hybrid question answering in relational algebra},
  author={Glenn, Parker and Dakle, Parag Pravin and Wang, Liang and Raghavan, Preethi},
  journal={arXiv preprint arXiv:2402.17882},
  year={2024}
}

@article{gao2023text,
author = {Gao, Dawei and Wang, Haibin and Li, Yaliang and Sun, Xiuyu and Qian, Yichen and Ding, Bolin and Zhou, Jingren},
title = {Text-to-SQL Empowered by Large Language Models: A Benchmark Evaluation},
year = {2024},
issue_date = {January 2024},
publisher = {VLDB Endowment},
volume = {17},
number = {5},
issn = {2150-8097},
url = {https://doi.org/10.14778/3641204.3641221},
doi = {10.14778/3641204.3641221},
journal = {Proc. VLDB Endow.},
month = jan,
pages = {1132–1145},
numpages = {14}
}

@inproceedings{zhao2023llm,
  title={Llm-sql-solver: Can llms determine SQL equivalence?},
  author={Zhao, Fuheng and Chen, Jiayue and Lim, Lawrence and Ahmad, Ishtiyaque and Agrawal, Divyakant and El Abbadi, Amr},
  booktitle={2025 IEEE International Conference on Big Data (BigData)},
  pages={1887--1894},
  year={2025},
  organization={IEEE}
}

@misc{google_bigquery_bqml,
  title        = {Introduction to AI and ML in BigQuery},
  author       = {{Google Cloud}},
  year         = {2025},
  month        = aug,
  note         = {Accessed: 2025-08-17},
  url          = {https://cloud.google.com/bigquery/docs/bqml-introduction}
}

@misc{snowflake_cortex_aisql,
  title        = {Snowflake Cortex AISQL (including LLM functions)},
  author       = {{Snowflake, Inc.}},
  year         = {2025},
  month        = jun,
  note         = {Preview feature documentation; accessed: 2025-08-17},
  url          = {https://docs.snowflake.com/user-guide/snowflake-cortex/aisql?lang=de/}
}

@misc{aws_bedrock_redshift,
  title        = {Bringing Generative AI to the Data Warehouse with Amazon Bedrock and Amazon Redshift},
  author       = {{Amazon Web Services}},
  year         = {2024},
  month        = feb,
  note         = {AWS re:Post article; accessed: 2025-08-17},
  url          = {https://repost.aws/articles/ARJszlMEepRti6xoM-0fsBmw/bringing-generative-ai-to-the-data-warehouse-with-amazon-bedrock-and-amazon-redshift}
}

@article{zhu2023large,
  title={Large language models for information retrieval: A survey},
  author={Zhu, Yutao and Yuan, Huaying and Wang, Shuting and Liu, Jiongnan and Liu, Wenhan and Deng, Chenlong and Chen, Haonan and Liu, Zheng and Dou, Zhicheng and Wen, Ji-Rong},
  journal={ACM Transactions on Information Systems},
  volume={44},
  number={1},
  pages={1--54},
  year={2025},
  publisher={ACM New York, NY}
}

@article{zhou2024serving,
  title={Serving Deep Learning Models from Relational Databases},
  author={Zhou, Lixi and Lin, Qi and Chowdhury, Kanchan and Masood, Saif and Eichenberger, Alexandre and Min, Hong and Sim, Alexander and Wang, Jie and Wang, Yida and Wu, Kesheng and others},
  journal={Advances in Database Technology-EDBT},
  volume={27},
  number={3},
  pages={717--724},
  year={2024}
}

@inproceedings{drozdov2023parade,
    title = "{P}a{R}a{D}e: Passage Ranking using Demonstrations with {LLM}s",
    author = "Drozdov, Andrew  and
      Zhuang, Honglei  and
      Dai, Zhuyun  and
      Qin, Zhen  and
      Rahimi, Razieh  and
      Wang, Xuanhui  and
      Alon, Dana  and
      Iyyer, Mohit  and
      McCallum, Andrew  and
      Metzler, Donald  and
      Hui, Kai",
    editor = "Bouamor, Houda  and
      Pino, Juan  and
      Bali, Kalika",
    booktitle = "Findings of the Association for Computational Linguistics: EMNLP 2023",
    month = dec,
    year = "2023",
    address = "Singapore",
    publisher = "Association for Computational Linguistics",
    url = "https://aclanthology.org/2023.findings-emnlp.950/",
    doi = "10.18653/v1/2023.findings-emnlp.950",
    pages = "14242--14252",
}

@inproceedings{sachan2022improving,
  title={Improving passage retrieval with zero-shot question generation},
  author={Sachan, Devendra and Lewis, Mike and Joshi, Mandar and Aghajanyan, Armen and Yih, Wen-tau and Pineau, Joelle and Zettlemoyer, Luke},
  booktitle={Proceedings of the 2022 Conference on Empirical Methods in Natural Language Processing},
  pages={3781--3797},
  year={2022}
}

@inproceedings{qin2023large,
  title={Large language models are effective text rankers with pairwise ranking prompting},
  author={Qin, Zhen and Jagerman, Rolf and Hui, Kai and Zhuang, Honglei and Wu, Junru and Yan, Le and Shen, Jiaming and Liu, Tianqi and Liu, Jialu and Metzler, Donald and others},
  booktitle={Findings of the Association for Computational Linguistics: NAACL 2024},
  pages={1504--1518},
  year={2024}
}

@inproceedings{luo2024prp,
  title={Prp-graph: Pairwise ranking prompting to llms with graph aggregation for effective text re-ranking},
  author={Luo, Jian and Chen, Xuanang and He, Ben and Sun, Le},
  booktitle={Proceedings of the 62nd Annual Meeting of the Association for Computational Linguistics (Volume 1: Long Papers)},
  pages={5766--5776},
  year={2024}
}

@article{shah2018simple,
  title={Simple, robust and optimal ranking from pairwise comparisons},
  author={Shah, Nihar B and Wainwright, Martin J},
  journal={Journal of machine learning research},
  volume={18},
  number={199},
  pages={1--38},
  year={2018}
}

@article{ma2023zero,
  title={Zero-shot listwise document reranking with a large language model},
  author={Ma, Xueguang and Zhang, Xinyu and Pradeep, Ronak and Lin, Jimmy},
  journal={arXiv preprint arXiv:2305.02156},
  year={2023}
}

@inproceedings{sun2023chatgpt,
  title={Is ChatGPT good at search? investigating large language models as re-ranking agents},
  author={Sun, Weiwei and Yan, Lingyong and Ma, Xinyu and Wang, Shuaiqiang and Ren, Pengjie and Chen, Zhumin and Yin, Dawei and Ren, Zhaochun},
  booktitle={Proceedings of the 2023 conference on empirical methods in natural language processing},
  pages={14918--14937},
  year={2023}
}

@article{pradeep2023rankvicuna,
  title={Rankvicuna: Zero-shot listwise document reranking with open-source large language models},
  author={Pradeep, Ronak and Sharifymoghaddam, Sahel and Lin, Jimmy},
  journal={arXiv preprint arXiv:2309.15088},
  year={2023}
}

@inproceedings{selinger1979access,
  title={Access path selection in a relational database management system},
  author={Selinger, P Griffiths and Astrahan, Morton M and Chamberlin, Donald D and Lorie, Raymond A and Price, Thomas G},
  booktitle={Proceedings of the 1979 ACM SIGMOD international conference on Management of data},
  pages={23--34},
  year={1979}
}

@inproceedings{alaparthi2025scalellm,
  title={ScaleLLM: A Technique for Scalable LLM-augmented Data Systems},
  author={Alaparthi, Ashwin and Loh, Paul and Marcus, Ryan},
  booktitle={Companion of the 2025 International Conference on Management of Data},
  pages={11--14},
  year={2025}
}

@misc{openintro_nba_heights,
  title        = {NBA Player Heights (2008--09 Season)},
  author       = {{OpenIntro}},
  howpublished = {R package \texttt{openintro}, dataset \texttt{nba\_heights}},
  year         = {2025},
  note         = {Available at \url{https://www.openintro.org/data/index.php?data=nba_heights}, accessed 2025-08-25},
}

@article{DL19,
  author       = {Nick Craswell and
                  Bhaskar Mitra and
                  Emine Yilmaz and
                  Daniel Campos and
                  Ellen M. Voorhees},
  title        = {Overview of the {TREC} 2019 deep learning track},
  journal      = {CoRR},
  volume       = {abs/2003.07820},
  year         = {2020},
  url          = {https://arxiv.org/abs/2003.07820},
  eprinttype    = {arXiv},
  eprint       = {2003.07820},
  bibsource    = {dblp computer science bibliography, https://dblp.org}
}

@article{wei2022chain,
  title={Chain-of-thought prompting elicits reasoning in large language models},
  author={Wei, Jason and Wang, Xuezhi and Schuurmans, Dale and Bosma, Maarten and Xia, Fei and Chi, Ed and Le, Quoc V and Zhou, Denny and others},
  journal={Advances in neural information processing systems},
  volume={35},
  pages={24824--24837},
  year={2022}
}

@misc{openai_structured_outputs,
  title        = {Structured model outputs},
  author       = {{OpenAI}},
  year         = {2024},
  howpublished = {OpenAI API Guide},
  note         = {Structured outputs ensure model responses adhere to a supplied JSON Schema. Accessed: 2025-08-25},
  url          = {https://platform.openai.com/docs/guides/structured-outputs}
}

@article{kendall1938new,
  title={A new measure of rank correlation},
  author={Kendall, Maurice G and others},
  journal={Biometrika},
  volume={30},
  number={1-2},
  pages={81--93},
  year={1938},
  publisher={Oxford University Press}
}

@inproceedings{wang2013theoretical,
  title={A theoretical analysis of NDCG type ranking measures},
  author={Wang, Yining and Wang, Liwei and Li, Yuanzhi and He, Di and Liu, Tie-Yan},
  booktitle={Conference on learning theory},
  pages={25--54},
  year={2013},
  organization={PMLR}
}

@inproceedings{lin2021pyserini,
  title     = {{Pyserini}: A {Python} Toolkit for Reproducible Information Retrieval Research with Sparse and Dense Representations},
  author    = {Lin, Jimmy and Ma, Xueguang and Lin, Sheng-Chieh and Yang, Jheng-Hong and Pradeep, Ronak and Nogueira, Rodrigo},
  booktitle = {Proceedings of the 44th International ACM SIGIR Conference on Research and Development in Information Retrieval (SIGIR)},
  year      = {2021},
}

@article{robertson2009probabilistic,
  title={The probabilistic relevance framework: BM25 and beyond},
  author={Robertson, Stephen and Zaragoza, Hugo and others},
  journal={Foundations and Trends{\textregistered} in Information Retrieval},
  volume={3},
  number={4},
  pages={333--389},
  year={2009},
  publisher={Now Publishers, Inc.}
}

@article{liang2022holistic,
  title={Holistic evaluation of language models},
  author={Liang, Percy and Bommasani, Rishi and Lee, Tony and Tsipras, Dimitris and Soylu, Dilara and Yasunaga, Michihiro and Zhang, Yian and Narayanan, Deepak and Wu, Yuhuai and Kumar, Ananya and others},
  journal={arXiv preprint arXiv:2211.09110},
  year={2022}
}

@article{patel2024lotus,
  title={Lotus: Enabling semantic queries with llms over tables of unstructured and structured data},
  author={Patel, Liana and Jha, Siddharth and Guestrin, Carlos and Zaharia, Matei},
  journal={arXiv e-prints},
  pages={arXiv--2407},
  year={2024}
}

@article{sun2025agenticdata,
  title={AgenticData: An Agentic Data Analytics System for Heterogeneous Data},
  author={Sun, Ji and Li, Guoliang and Zhou, Peiyao and Ma, Yihui and Xu, Jingzhe and Li, Yuan},
  journal={arXiv preprint arXiv:2508.05002},
  year={2025}
}

@article{gong2025sqlens,
  title={Sqlens: An end-to-end framework for error detection and correction in text-to-sql},
  author={Gong, Yue and Lei, Chuan and Qin, Xiao and Vaidya, Kapil and Narayanaswamy, Balakrishnan and Kraska, Tim},
  journal={Advances in Neural Information Processing Systems},
  volume={38},
  pages={135571--135604},
  year={2026}
}

@inproceedings{nogueira2020documentrankingpretrainedsequencetosequence,
  title={Document ranking with a pretrained sequence-to-sequence model},
  author={Nogueira, Rodrigo and Jiang, Zhiying and Pradeep, Ronak and Lin, Jimmy},
  booktitle={Findings of the association for computational linguistics: EMNLP 2020},
  pages={708--718},
  year={2020}
}

@inproceedings{Martinez2004partial,
  title={Partial quicksort},
  author={Mart{\i}nez, Conrado},
  booktitle={Proc. 6th ACMSIAM Workshop on Algorithm Engineering and Experiments and 1st ACM-SIAM Workshop on Analytic Algorithmics and Combinatorics},
  pages={224--228},
  year={2004}
}

@inproceedings{wu2016samplingbasedqueryreoptimization,
  title={Sampling-based query re-optimization},
  author={Wu, Wentao and Naughton, Jeffrey F and Singh, Harneet},
  booktitle={Proceedings of the 2016 international conference on management of data},
  pages={1721--1736},
  year={2016}
}

@inproceedings{Gibbons1998,
  title={New sampling-based summary statistics for improving approximate query answers},
  author={Gibbons, Phillip B and Matias, Yossi},
  booktitle={Proceedings of the 1998 ACM SIGMOD international conference on Management of data},
  pages={331--342},
  year={1998}
}

@inproceedings{Chen2006,
  title={Two-level sampling for join size estimation},
  author={Chen, Yu and Yi, Ke},
  booktitle={Proceedings of the 2017 ACM International Conference on Management of Data},
  pages={759--774},
  year={2017}
}

@article{gu2025surveyllmasajudge,
  title={A survey on llm-as-a-judge},
  author={Gu, Jiawei and Jiang, Xuhui and Shi, Zhichao and Tan, Hexiang and Zhai, Xuehao and Xu, Chengjin and Li, Wei and Shen, Yinghan and Ma, Shengjie and Liu, Honghao and others},
  journal={The Innovation},
  volume={7},
  number={6},
  year={2026},
  publisher={Elsevier}
}

@inproceedings{kim2024flexexpertlevelfalselessexecution,
  title={Flex: Expert-level false-less execution metric for text-to-sql benchmark},
  author={Kim, Heegyu and Taeyang, Jeon and Choi, Seunghwan and Choi, Seungtaek and Cho, Hyunsouk},
  booktitle={Proceedings of the 2025 Conference of the Nations of the Americas Chapter of the Association for Computational Linguistics: Human Language Technologies (Volume 1: Long Papers)},
  pages={4448--4475},
  year={2025}
}

@article{Emerson2013Borda,
  title={The original Borda count and partial voting},
  author={Emerson, Peter},
  journal={Social Choice and Welfare},
  volume={40},
  number={2},
  pages={353--358},
  year={2013},
  publisher={Springer}
}

@inproceedings{cortexaisql,
  title={Cortex aisql: A production sql engine for unstructured data},
  author={Liskowski, Pawe{\l} and Han, Benjamin and Aggarwal, Paritosh and Chen, Bowei and Jiang, Boxin and Jindal, Nitish and Li, Zihan and Lin, Aaron and Schmaus, Kyle and Tayade, Jay and others},
  booktitle={Companion of the International Conference on Management of Data},
  pages={400--412},
  year={2026}
}

@article{morris2025languagemodelsmemorize,
  title={How much do language models memorize?},
  author={Morris, John X and Sitawarin, Chawin and Guo, Chuan and Kokhlikyan, Narine and Suh, G Edward and Rush, Alexander M and Chaudhuri, Kamalika and Mahloujifar, Saeed},
  journal={arXiv preprint arXiv:2505.24832},
  year={2025}
}

@inproceedings{Brown_2021,
  title={When is memorization of irrelevant training data necessary for high-accuracy learning?},
  author={Brown, Gavin and Bun, Mark and Feldman, Vitaly and Smith, Adam and Talwar, Kunal},
  booktitle={Proceedings of the 53rd annual ACM SIGACT symposium on theory of computing},
  pages={123--132},
  year={2021}
}

@inproceedings{wang2025generalizationvsmemorizationtracing,
  title={Generalization vs memorization: Tracing language models’ capabilities back to pretraining data},
  author={Wang, Xinyi and Antoniades, Antonis and Elazar, Yanai and Amayuelas, Alfonso and Albalak, Alon and Zhang, Kexun and Wang, William},
  booktitle={International Conference on Learning Representations},
  volume={2025},
  pages={49948--49968},
  year={2025}
}

@inproceedings{Wen2024Membership,
  title={Membership inference attacks against in-context learning},
  author={Wen, Rui and Li, Zheng and Backes, Michael and Zhang, Yang},
  booktitle={Proceedings of the 2024 on ACM SIGSAC Conference on Computer and Communications Security},
  pages={3481--3495},
  year={2024}
}

@article{Saari2023SelectingBorda,
  title={Selecting a voting method: the case for the Borda count},
  author={Saari, Donald G},
  journal={Constitutional Political Economy},
  volume={34},
  number={3},
  pages={357--366},
  year={2023},
  publisher={Springer}
}

@article{heilman2022noisestabilityrankedchoice,
  title={Noise Stability of Ranked Choice Voting},
  author={Heilman, Steven},
  journal={arXiv preprint arXiv:2209.11183},
  year={2022}
}

@misc{tanuprabhu2020population,
  author = {Tanu Prabhu},
  title = {Population by Country — 2020},
  howpublished = {\url{https://www.kaggle.com/datasets/tanuprabhu/population-by-country-2020}},
  year = {2020},
  note = {Accessed: 2025-11-24}
}

@inproceedings{Craswell2021TRECDL2020,
  title        = {TREC Deep Learning Track: Reusable Test Collections in the Large Data Regime},
  author       = {Nick Craswell and Bhaskar Mitra and Emine Yilmaz and Daniel Campos and Ellen M. Voorhees and Ian Soboroff},
  booktitle    = {Proceedings of the 44th International ACM SIGIR Conference on Research and Development in Information Retrieval (SIGIR ’21)},
  year         = {2021},
  pages        = {2369--2375},
  publisher    = {ACM},
  doi          = {10.1145/3404835.3463249},
  url          = {https://doi.org/10.1145/3404835.3463249}
}

@inproceedings{cheng2023batchpromptingefficientinference,
  title={Batch prompting: Efficient inference with large language model apis},
  author={Cheng, Zhoujun and Kasai, Jungo and Yu, Tao},
  booktitle={Proceedings of the 2023 Conference on Empirical Methods in Natural Language Processing: Industry Track},
  pages={792--810},
  year={2023}
}

@article{lao2025sembenchbenchmarksemanticquery,
author = {Lao, Jiale and Zimmerer, Andreas and Ovcharenko, Olga and Cong, Tianji and Russo, Matthew and Vitagliano, Gerardo and Cochez, Michael and {\"O}zcan, Fatma and Gupta, Gautam and Hottelier, Thibaud and Jagadish, H. V. and Kissel, Kris and Schelter, Sebastian and Kipf, Andreas and Trummer, Immanuel},
title = {SemBench: A Benchmark for Semantic Query Processing Engines},
year = {2026},
issue_date = {April 2026},
publisher = {VLDB Endowment},
volume = {19},
number = {8},
issn = {2150-8097},
url = {https://doi.org/10.14778/3811243.3811249},
doi = {10.14778/3811243.3811249},
journal = {Proc. VLDB Endow.},
month = jul,
pages = {1754–1767},
numpages = {14}
}

@misc{trulens,
  title        = {TruLens: Evals and Tracing for LLMs and Agents},
  howpublished = {\url{https://www.trulens.org/}},
  note         = {Accessed: 2025-11-25}
}

@article{wang2025unify,
  title = {Unify: A System For Unstructured Data Analytics},
  author = {Wang, Jiayi and Li, Yuan and Wu, Jianming and Xu, Shihui and Li, Guoliang},
  journal = {Proceedings of the VLDB Endowment},
  volume = {18},
  number = {12},
  pages = {5287--5290},
  year = {2025},
  doi = {10.14778/3750601.3750653}
}

@misc{BerriAI_litellm_2025,
  author       = {BerriAI},
  title        = {LiteLLM: Python SDK and proxy server for calling 100+ LLM APIs},
  howpublished = {GitHub repository},
  year         = {2025},
  url          = {https://github.com/BerriAI/litellm},
  note         = {Accessed: 2025-11-28}
}

@article{trummer2025implementingsemanticjoinoperators,
  title={Implementing semantic join operators efficiently},
  author={Trummer, Immanuel},
  journal={arXiv preprint arXiv:2510.08489},
  year={2025}
}

@article{knuth1997taocp,
  title={The art of computer programming, vol. 3: Searching and sorting},
  author={Knuth, Donald E},
  journal={Reading MA: Addison-Wisley},
  pages={543--583},
  year={1973}
}

@misc{openai_faq_temperature,
  title        = {OpenAI FAQ: How should I set the temperature parameter?},
  author       = {OpenAI},
  year         = {2024},
  howpublished = {\url{https://platform.openai.com/docs/faq/faq}},
  note         = {Accessed: 2025-11-30}
}

@inproceedings{Zhuang2024BeyondYesAndNo,
  title={Beyond yes and no: Improving zero-shot llm rankers via scoring fine-grained relevance labels},
  author={Zhuang, Honglei and Qin, Zhen and Hui, Kai and Wu, Junru and Yan, Le and Wang, Xuanhui and Bendersky, Michael},
  booktitle={Proceedings of the 2024 conference of the North American chapter of the Association for Computational Linguistics: Human language technologies (volume 2: short papers)},
  pages={358--370},
  year={2024}
}

@article{lin2025twixautomaticallyreconstructingstructured,
  title={Twix: Automatically reconstructing structured data from templatized documents},
  author={Lin, Yiming and Hasan, Mawil and Kosalge, Rohan and Cheung, Alvin and Parameswaran, Aditya G},
  journal={arXiv e-prints},
  pages={arXiv--2501},
  year={2025}
}

@inproceedings{papadopoulos2025haides,
  title        = {HAIDES: Adaptive Approximation of Inference Queries over Unstructured Data},
  author       = {Papadopoulos, Christos Chrysovalantis and Simitsis, Alkis and Pedersen, Torben Bach},
  booktitle    = {Proceedings of the 41st IEEE International Conference on Data Engineering (ICDE)},
  year         = {2025},
  pages        = {2394--2407}
}

@inproceedings{wang2025cequel,
  title        = {Cequel: Cost-Effective Querying of Large Language Models for Text Clustering},
  author       = {Wang, Hongtao and Zhang, Taiyan and Yang, Renchi and Xu, Jianliang},
  booktitle    = {Proceedings of the 34th ACM International Conference on Information and Knowledge Management (CIKM)},
  year         = {2025},
  pages        = {2998--3008},
  publisher    = {Association for Computing Machinery},
  doi          = {10.1145/3746252.3761074}
}

@article{sun2025quest,
  title={Quest: Query optimization in unstructured document analysis},
  author={Sun, Zhaoze and Deng, Qiyan and Chai, Chengliang and Jin, Kaisen and Guo, Xinyu and Han, Han and Yuan, Ye and Wang, Guoren and Cao, Lei},
  journal={arXiv preprint arXiv:2507.06515},
  year={2025}
}

@article{zhao2025neurdbdesignimplementationaipowered,
  title={Neurdb: On the design and implementation of an ai-powered autonomous database},
  author={Zhao, Zhanhao and Cai, Shaofeng and Gao, Haotian and Pan, Hexiang and Xiang, Siqi and Xing, Naili and Chen, Gang and Ooi, Beng Chin and Shen, Yanyan and Wu, Yuncheng and others},
  journal={arXiv preprint arXiv:2408.03013},
  year={2024}
}

@article{ji2025table,
  title={Table integration in data lakes unleashed: pairwise integrability judgment, integrable set discovery, and multi-tuple conflict resolution},
  author={Ji, Daomin and Luo, Hui and Bao, Zhifeng and Culpepper, J Shane},
  journal={The VLDB Journal},
  volume={34},
  number={3},
  pages={36},
  year={2025},
  publisher={Springer}
}

@article{noroozizadeh2025deepsequencemodelstend,
  title={Deep sequence models tend to memorize geometrically; it is unclear why},
  author={Noroozizadeh, Shahriar and Nagarajan, Vaishnavh and Rosenfeld, Elan and Kumar, Sanjiv},
  journal={arXiv preprint arXiv:2510.26745},
  year={2025}
}

@article{wang2023selfconsistencyimproveschainthought,
  title={Self-consistency improves chain of thought reasoning in language models},
  author={Wang, Xuezhi and Wei, Jason and Schuurmans, Dale and Le, Quoc and Chi, Ed and Narang, Sharan and Chowdhery, Aakanksha and Zhou, Denny},
  journal={arXiv preprint arXiv:2203.11171},
  year={2022}
}

@inproceedings{RRF,
author = {Cormack, Gordon V. and Clarke, Charles L A and Buettcher, Stefan},
title = {Reciprocal rank fusion outperforms condorcet and individual rank learning methods},
year = {2009},
isbn = {9781605584836},
publisher = {Association for Computing Machinery},
address = {New York, NY, USA},
url = {https://doi.org/10.1145/1571941.1572114},
doi = {10.1145/1571941.1572114},
series = {SIGIR '09}
}

@inproceedings{zellers2019hellaswag,
  title={Hellaswag: Can a machine really finish your sentence?},
  author={Zellers, Rowan and Holtzman, Ari and Bisk, Yonatan and Farhadi, Ali and Choi, Yejin},
  booktitle={Proceedings of the 57th annual meeting of the association for computational linguistics},
  pages={4791--4800},
  year={2019}
}

@inproceedings{boteva2016full,
  title={A full-text learning to rank dataset for medical information retrieval},
  author={Boteva, Vera and Gholipour, Demian and Sokolov, Artem and Riezler, Stefan},
  booktitle={European Conference on Information Retrieval},
  pages={716--722},
  year={2016},
  organization={Springer}
}

@article{googleproxy,
  title={100x Cost \& Latency Reduction: Performance Analysis of AI Query Approximation using Lightweight Proxy Models:[Experiments \& Analysis]},
  author={Chung, Yeounoh and Desai, Rushabh and He, Jian and Xiao, Yu and Hottelier, Thibaud and Kom Samo, Yves-Laurent and Khadilkar, Pushkar and Chen, Xianshun and Idicula, Sam and Ozcan, Fatma and others},
  journal={Proceedings of the ACM on Management of Data},
  volume={4},
  number={3 (SIGMOD},
  pages={1--23},
  year={2026},
  publisher={ACM, New York, NY, USA}
}

@article{LiskowskiCompositionalOL,
  title={Compositional Online Learning for Semantic Data Processing Systems},
  author={Liskowski, Pawe{\l} and Zhao, Fuheng and Han, Benjamin and Datta, Anupam and Tsirogiannis, Dimitris},
  journal={arXiv preprint arXiv:2608.27244},
  year={2026}
}

@inproceedings{liu2025leveraging,
  title={Leveraging passage embeddings for efficient listwise reranking with large language models},
  author={Liu, Qi and Wang, Bo and Wang, Nan and Mao, Jiaxin},
  booktitle={Proceedings of the ACM on Web Conference 2025},
  pages={4274--4283},
  year={2025}
}

@article{kwiatkowski2019natural,
  title={Natural questions: a benchmark for question answering research},
  author={Kwiatkowski, Tom and Palomaki, Jennimaria and Redfield, Olivia and Collins, Michael and Parikh, Ankur and Alberti, Chris and Epstein, Danielle and Polosukhin, Illia and Devlin, Jacob and Lee, Kenton and others},
  journal={Transactions of the Association for Computational Linguistics},
  volume={7},
  pages={453--466},
  year={2019},
  publisher={MIT Press, One Rogers Street, Cambridge, MA 02142-1209, USA journals-info~…}
}

@inproceedings{zhang-choi-2021-situatedqa,
    title = "{S}ituated{QA}: Incorporating Extra-Linguistic Contexts into {QA}",
    author = "Zhang, Michael  and
      Choi, Eunsol",
    editor = "Moens, Marie-Francine  and
      Huang, Xuanjing  and
      Specia, Lucia  and
      Yih, Scott Wen-tau",
    booktitle = "Proceedings of the 2021 Conference on Empirical Methods in Natural Language Processing",
    month = nov,
    year = "2021",
    address = "Online and Punta Cana, Dominican Republic",
    publisher = "Association for Computational Linguistics",
    url = "https://aclanthology.org/2021.emnlp-main.586/",
    doi = "10.18653/v1/2021.emnlp-main.586",
    pages = "7371--7387",
}

@String{Computing = "Computing" }

@String{Computer = "{IEEE} Computer" }

@String{Springer = "Springer-Verlag" }

@ArtifactSoftware{R,
    title = {R: A Language and Environment for Statistical Computing},
    author = {{R Core Team}},
    organization = {R Foundation for Statistical Computing},
    address = {Vienna, Austria},
    year = {2019},
    url = {https://www.R-project.org/},
}


\end{document}